%% file: main.tex
\documentclass{aastex63}
\usepackage[T1]{fontenc}
\usepackage{color}
\usepackage{amsmath}
\usepackage{natbib}
\usepackage{ctable}
\usepackage{bm}
\usepackage{xspace}

\usepackage{graphicx}

\newcommand{\given}{\,|\,}

\newcommand{\mpc}{{\rm Mpc}}
\newcommand{\eg}{\emph{e.g.}}
\newcommand{\ie}{\emph{i.e.}}

\let\oldAA\AA
\renewcommand{\AA}{\text{\normalfont\oldAA}}
\newcommand{\gr}{g{-}r}
\newcommand{\fnuv}{FUV{-}NUV}
\newcommand{\sfr}{{\rm SFR}}
\newcommand{\ssfr}{{\rm SSFR}}
\newcommand{\mtaum}{m_{\tau,M_*}}
\newcommand{\mtaus}{m_{\tau,{\rm SSFR}}}

\newcommand{\mdeltam}{m_{\delta,M_*}}
\newcommand{\mdeltas}{m_{\delta,{\rm SFR}}}

\newcommand{\eda}{EDA}

\newcommand{\foreign}[1]{\textsl{#1}}
\newcommand{\etal}{\foreign{et~al.}}

\newcommand{\bitem}{\begin{itemize}}
\newcommand{\eitem}{\end{itemize}}
\newcommand{\beq}{\begin{equation}}
\newcommand{\eeq}{\end{equation}}

\shorttitle{The Empirical Dust Attenuation Framework}
\shortauthors{Hahn et al.}

\begin{document} 

\title{IQ Collaboratory III: The Empirical Dust Attenuation Framework ---
Taking Hydrodynamical Simulations with a Grain of Dust}

\author[0000-0003-1197-0902]{ChangHoon Hahn}
\correspondingauthor{ChangHoon Hahn}
\email{changhoon.hahn@princeton.edu}
\affil{Department of Astrophysical Sciences, Princeton University, Peyton Hall, Princeton NJ 08544, USA} 

\author{Tjitske K. Starkenburg}
\affil{Center for Interdisciplinary Exploration and Research in Astrophysics (CIERA) and \\Department of Physics and Astronomy, 1800 Sherman Ave, Evanston IL 60201, USA}

\author{Daniel Angl\'es-Alc\'azar}
\affil{Department of Physics, University of Connecticut, 196 Auditorium Road,
U-3046, Storrs, CT 06269-3046, USA}
\affil{Center for Computational Astrophysics, Flatiron Institute, 162 5th Ave,
New York, NY 10010, USA}

\author[0000-0002-8131-6378]{Ena Choi}
\affil{Quantum Universe Center, 
        Korea Institute for Advanced Study, 
        Hoegiro 85, Seoul 02455, Korea}

\author{Romeel Dav\'e}
\affil{Institute for Astronomy, Royal Observatory, Univ. of Edinburgh, Edinburgh EH9 3HJ, UK}
\affil{University of the Western Cape, Bellville, Cape Town 7535, South Africa}
\affil{South African Astronomical Observatories, Observatory, Cape Town 7925, South Africa}

\author[0000-0002-1081-3991]{Claire Dickey} 
\affil{Department of Astronomy, Yale University, New Haven CT 06520 USA}

\author[0000-0001-9298-3523]{Kartheik G. Iyer}
\affil{Dunlap Institute for Astronomy and Astrophysics, University of Toronto,
50 St George St, Toronto, ON M5S 3H4, Canada}

\author[0000-0003-2060-8331]{Ariyeh H. Maller}
\affil{Department of Physics, New York City College of Technology, City
University of New York, 300 Jay St., Brooklyn, NY 11201, USA}

\author{Rachel S. Somerville}
\affil{Department of Physics and Astronomy, Rutgers University, 136
Frelinghuysen Road, Piscataway, NJ 08854, USA}
\affil{Center for Computational Astrophysics, Flatiron Institute, 162 5th Ave,
New York, NY 10010, USA}

\author{Jeremy L. Tinker}
\affil{Center for Cosmology and Particle Physics, Department of Physics, New York University, 4 Washington Place, New York, NY 10003}

\author{L. Y. Aaron Yung}
\affil{Astrophysics Science Division, NASA Goddard Space Flight Center, 8800 Greenbelt Rd, Greenbelt, MD 20771, USA}



\begin{abstract}
    We present the Empirical Dust Attenuation (\eda) framework --- a flexible
    prescription for 
    assigning realistic dust attenuation to simulated galaxies based on their
    physical properties. 
    We use the \eda~to forward model synthetic observations for three state-of-the-art large-scale cosmological
    hydrodynamical simulations: SIMBA, IllustrisTNG, and EAGLE. 
    We then compare the optical
    and UV color-magnitude relations, $(\gr) - M_r$ and $(\fnuv)-M_r$, of the
    simulations to a $M_r < -20$ and UV complete SDSS galaxy sample using
    likelihood-free inference. 
    Without dust, none of the simulations match observations, as expected.
    With the \eda, however, we can reproduce the observed color-magnitude
    with all three simulations. 
    Furthermore, the attenuation curves predicted by our dust prescription are
    in good agreement with the observed attenuation--slope relations and
    attenuation curves of star-forming galaxies. 
    However, the \eda~does not predict star-forming galaxies with low
    $A_V$ since simulated star-forming galaxies are intrinsically much
    brighter than observations.
    Additionally, the \eda~provides, for the first time, predictions
    on the attenuation curves of quiescent galaxies, which are challenging to
    measure observationally. 
    Simulated quiescent galaxies require shallower attenuation curves
    with lower amplitude than star-forming galaxies.
    The \eda, combined with forward modeling, provides an effective
    approach for shedding light on  dust in galaxies and
    probing hydrodynamical simulations. 
    This work also illustrates a major limitation in comparing galaxy
    formation models: by adjusting dust attenuation, simulations that predict
    significantly different galaxy populations can reproduce the same UV
    and optical observations. 
\end{abstract}

\keywords{
    galaxies: formation -- galaxies: evolution -- galaxies: statistics -- methods: numerical 
}
\NewPageAfterKeywords

\input{intro}
\input{sims} 
\input{dem}

\input{abc}
\input{results}
\input{summary}

\section*{Acknowledgements}
It's a pleasure to thank 
    Michael Blanton, 
    Nicholas T. Faucher, 
    Marla Geha, 
    Shy Genel,
    Jenny E. Green,
    Daniel Kelson, 
    Mariska Kriek, 
    Peter Melchior, 
    Desika Narayanan, 
    Samir Salim, 
    and Katherine Suess for
valuable discussions and comments.
This material is based upon work supported by the U.S. Department of Energy,
Office of Science, Office of High Energy Physics, under contract No.
DE-AC02-05CH11231. 
CH is supported by the AI Accelerator program of the Schmidt Futures Foundation.

We thank the Illustris collaboration and the Virgo Consortium for making their
simulation data publicly available, and the SIMBA collaboration for sharing
their data with us.
The EAGLE and SIMBA simulations were performed using the DiRAC-2 facility at
Durham, managed by the ICC, and the PRACE facility Curie based in France at
TGCC, CEA, Bruy\`{e}res-le-Ch\^{a}tel.

This research was supported in part through the computational resources and
staff contributions provided by the Quest high performance computing facility
at Northwestern University, which is jointly supported by the Office of the
Provost, the Office for Research, and Northwestern University Information
Technology. 

The data used in this work were, in part, hosted on facilities supported by the
Scientific Computing Core at the Flatiron Institute, a division of the Simons
Foundation, and the analysis was largely done using those facilities.
The IQ~(Isolated \& Quiescent)~Collaboratory thanks the Flatiron Institute for hosting the collaboratory and its meetings. 
The Flatiron Institute is supported by the Simons Foundation.
Funding for the Sloan Digital Sky Survey IV has been provided by the Alfred P.
Sloan Foundation, the U.S. Department of Energy Office of Science, and the
Participating Institutions. SDSS acknowledges support and resources from the
Center for High-Performance Computing at the University of Utah. The SDSS web
site is www.sdss.org.
The SDSS is managed by the Astrophysical Research Consortium for the
Participating Institutions.

\appendix
\input{slab.tex}

\bibliographystyle{mnras}
\bibliography{galpopfm} 
\end{document}

%% file: intro.tex
\section{Introduction} \label{sec:intro} 
Dust in the interstellar medium of a galaxy can dramatically impact its
spectral energy distribution (SED). 
The combined effect of dust on a galaxy's SED is typically described using
an attenuation curve, $A(\lambda)$, which has now been broadly characterized 
by observations. 
In UV, attenuation curves steeply rise due to absorption by small grains.
At $2175\AA$, in the near-UV (NUV), there is an absorption bump referred to as
the ``UV dust bump''. 
At longer optical wavelengths, the curves take on a power-law shape. 
Finally, dust reemits the light attenuated in the optical and UV in the
infrared~\citep[for an overview see][]{calzetti2001, draine2003,
galliano2018}.
By impacting the SED, dust also affects the physical properties of a galaxy
that are inferred from the SED, such as its star formation rate ($\sfr$),
stellar mass ($M_*$), or star formation history~\citep[see reviews by][]{walcher2011,
conroy2013}. 
Assumptions on dust attenuation can dramatically vary these
properties~\citep{kriek2013, reddy2015, salim2016, salim2020}.
Since these properties are the building blocks to our understanding of
galaxies and how they evolve, a better understanding of dust not only provides
insights into dust, but also underpins all galaxy studies.  

To better understand dust in galaxies, many observational works have
examined trends between dust attenuation and galaxy properties.
For example, UV and optical attenuation are found to correlate with galaxy
properties such as $M_*$, $\sfr$, and metallicity in star-forming
galaxies~\citep[][for a recent review see \citealt{salim2020}]{garn2010, battisti2016}. 
The slope of the attenuation curves in star-forming galaxies also correlate
with galaxy properties, such as $M_*$, specific $\sfr$ ($\ssfr$), metallicity,
and axial ratio~\citep{wild2011, battisti2017}. 
Despite the progress, there is still no clear consensus on the connection
between dust attenuation and galaxy properties.
Furthermore, studies so far have focused mainly on star-forming galaxies and
little is known about dust attenuation in quiescent galaxies. 
A major limitation of observational approaches is that dust attenation and
galaxy properties measured from galaxy SEDs are model-dependent and subject to
variations, inconsistencies, and biases of different methodologies.
Different methods can measure vastly different derived values, even for the
same observations~\citep[\eg][see also Appendix~\ref{sec:slab}]{speagle2014, katsianis2020}.
SED modeling can also impose undesirable priors on derived galaxy
properties~\citep{carnall2018, leja2019} and suffer from parameter degeneracies
that are poorly understood. 


Significant progress has also been made in theoretically modeling dust. 
Simulations can now model the radiative transfer of stellar light
through a dusty ISM for a wide range of configurations:
from simple slab-like dust geometries~\citep[\eg][]{witt1996, witt2000,
seon2016} to 3D hydrodynamical simulations of entire
galaxies~\citep[\eg][]{jonsson2006, rocha2008, hayward2015, natale2015,
hou2017}. 
Radiative transfer models have even been applied to cosmological
hydrodynamical simulations~\cite[\eg][]{camps2015, narayanan2018,
cochrane2019, rodriguez-gomez2019, trayford2020}. 
Dust has also been examined in a cosmological context using 
semi-analytic models ~\citep[SAMs; \eg][]{granato2000, fontanot2009, wilkins2012,
gonzalez-perez2013, popping2017}. 
Yet there are still major limitations in modeling dust. 
Dust models in cosmological simulations currently do not reproduce the
redshift evolution of dust properties~(\citealp{somerville2012, yung2019,
vogelsberger2020}, but see \citealp{li2019}) and radiative transfer models
produce attenuation-slope relations that are significantly steeper than observations. 
Many models also require significant hand-tuning (\eg~propagating rays/photons into
particular cells) and make assumptions on the underlying dust grain models~\citep[see][for a review]{steinacker2013}. 
Lastly, radiative transfer models are computationally expensive.  
Applying a range of radiative transfer dust models to multiple simulations for
comparisons would require huge computational resources.
Using them with Monte Carlo sampling methods for parameter exploration or to
marginalize over the impact of dust would be prohibitive.  

We take a different approach from the observational and theoretical works
above --- \emph{we investigate dust attenuation using a forward modeling
approach to compare simulations to observations}.
Our ``forward model'' starts with three major large-scale hydrodynamical
simulations: EAGLE \citep{schaye2015}, IllustrisTNG \citep{nelson2019},
and SIMBA \citep{dave2019}. 
We use their outputs (\eg~star formation history) to build SEDs for each
simulated galaxy.
We then apply dust attenuation to the SEDs using the Empirical Dust
Attenuation (\eda) framework, which we describe shortly.
We construct photometry from the attenuated SEDs and afterwards apply a 
realistic noise model and sample selection function to construct synthetic
observations. 
Finally, we compare the synthetic observations to actual observations and
constrain the free parameters of our EDA prescription. 
The EDA parameter constraints then provide insight into dust attenation and its
connection to galaxies.

The Empirical Dust Attenuation framework provides a flexible and
computationally inexpensive prescription for statistically assigning
attenatuion curves to simulated galaxy based on their physical properties. 
In this work, we assign attenuation curves based on $M_*$, $\ssfr$, and the
\eda~parameters, as well as randomly sampled galaxy inclinations that
introduces stochasticity.  
The \eda~parameters set the strength of the $M_*$ and $\ssfr$ dependence in the
amplitude and slope of the attenuation curves. 
Unlike radiative transfer models, the \eda~does not seek to produce realistic
dust attenuation for individual galaxies. 
However, as we later demonstrate, it produces realistic distributions of dust
attenuation for galaxy populations. 
The \eda~provides an empirical framework for dust attenuation, analogous to the
halo occupation framework in galaxy formation~\citep[for a review see ][]{wechsler2018}. 

With a forward modeling approach, we compare simulations to  observations
entirely in observational space, so we are not impacted by
the inconsistencies of observational methods for measuring galaxy properties. 
Forward models can also directly include the selection functions and
observational systematic effects so they can account for these effects to
exploit the full observational data set.
Furthermore, since we use the \eda~in our forward model, we can apply a
wide range of realistic dust attenuation curves to simulated galaxies in a
matter of seconds, unlike radiative transfer models.
This means we can easily explore and sample the dust parameter space and infer
the relationship between dust attenuation and galaxy properties. 
That is the focus of this paper. 
Beyond investigating dust, the \eda~also provides a framework where we can
treat dust as {\em nuisance} parameters and tractably marginalize over dust
attenuation. 
In the subsequent paper of the IQ series, Starkenburg et al. (in
preparation), we will use the \eda~framework to compare star formation
quenching in cosmological galaxy formation models after marginalizing over
dust attenuation. 


In Section~\ref{sec:sims}, we describe the three large-scale cosmological
hydrodynamical simulations (SIMBA, IllustrisTNG, and EAGLE) that we use in
our forward model along with the observed SDSS galaxy sample used for
comparison. 
Next, we present the \eda~prescription used in this work
(Section~\ref{sec:dem}) and the likelihood-free inference method used to 
compare the simulations to observations (Section~\ref{sec:abc}). 
Finally, in Section~\ref{sec:results}, we present the results of our
comparison and discuss their implications on dust attenuation and its
connection to galaxy properties. 

%% file: sims.tex
\begin{figure}
\begin{center}
    \includegraphics[width=\textwidth]{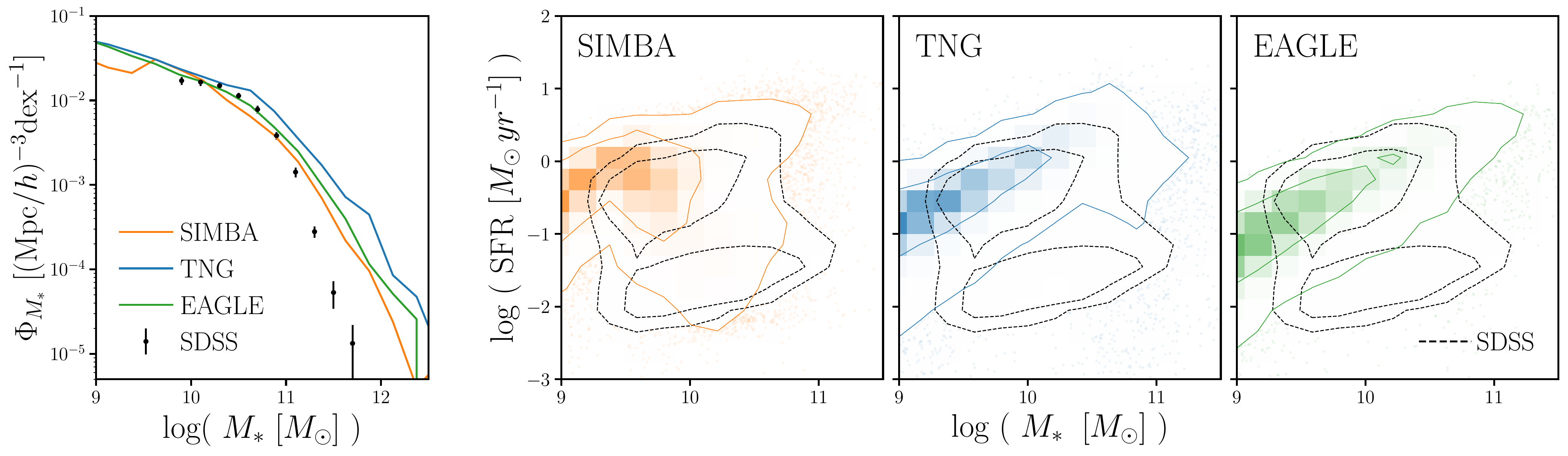}
    \caption{\label{fig:smf_msfr}
    The stellar mass functions, $\Phi_{M_*}$ (left-most panel), and $M_*-\sfr$
    relation (right panels) of galaxies in three cosmological hydrodynamic 
    simulations: SIMBA (orange), TNG (blue), and EAGLE (green). 
    For reference, we include $\Phi_{M_*}$ and the $M_*-\sfr$ relation of
    SDSS observations (black).
    Uncertainties for the SDSS $\Phi_{M_*}$ are derived using jackknife
    resampling. 
    We describe the simulations and observations in Section~\ref{sec:sims}. 
    Although, we see significant differences between the $M_*-\sfr$ relations of
    SDSS and the simulations, the SDSS measurements cannot be directly compared
    to the simulations' theoretical $M_*$ and $\sfr$ predictions.
    A forward modeling approach, where we construct synthetic observations
    for the simulations, enables an apples-to-apples comparison between
    simulations and observation. 
    \emph{Furthermore, differences
    in $\Phi_{M_*}$ and the $M_*-\sfr$ relations among the hydrodynamic simulations 
    highlight how they predict galaxy populations with significantly different
    physical properties.} 
    }
\end{center}
\end{figure}

\section{Data}\label{sec:sims}
In this paper, we present the Empirical Dust Attenuation (\eda) model and
demonstrate how it can be used in a forward modeling approach to compare galaxy
populations in simulations and observations. For our simulations, we use three large-scale
cosmological hydrodynamical simulations: the IllustrisTNG (hereafter TNG),
EAGLE, and SIMBA. For our observations, we use a galaxy sample derived from
SDSS. Below, we briefly describe the simulations and the SDSS observations used
throughout this work.

In Figure~\ref{fig:smf_msfr}, we present the stellar mass functions,
$\Phi_{M_*}$ (left-most panel), and $M_*-\sfr$ relations (right panels) of
galaxies in SIMBA (orange), TNG (blue), and EAGLE (green). For reference, we
include $\Phi_{M_*}$ and the 
$M_*-\sfr$ relation for SDSS observations. For the simulations, $M_*$ is the
total stellar mass within the subhalo and $\sfr$ is the instantaneous $\sfr$ in the
dense and cold star-forming gas. 
We do not impose any selection cuts on the simulations in Figure~\ref{fig:smf_msfr}. 
For SDSS, we use $\mathtt{kcorrect}$ $M_*$ assuming a~\cite{chabrier2003}
initial mass function~\citep{blanton2007a} and $\sfr$ from the
current release of \cite{brinchmann2004}\footnote{\url{http://www.mpa-garching.mpg.de/SDSS/DR7/}}.
The uncertainties for the SDSS SMF are derived from jackknife resampling.
Although, we find striking differences between the $M_*-\sfr$ relation of SDSS
and the simulations, the $M_*$ and $\sfr$ of simulations are theoretical
predictions while the SDSS values are measurements, which are subject to, for
instance, $\sfr$ measurement limits, inconsistencies among $\sfr$ tracers, and
aperture effects.
This inconsistency prevents a meaningful direct comparison. 
However, with a forward modeling approach of constructing synthetic
observations, the simulations can be directly compared to
observations~\citep[see \eg][Starkenburg et al. in prep.]{dickey2020}.
Figure~\ref{fig:smf_msfr} also illustrates that the hydrodynamical simulations
predict significantly different SMFs and $M_*-\sfr$ relations from each other.
This difference, which was also recently highlighted in \cite{hahn2019c},
demonstrates that \emph{the hydrodynamical simulations predict galaxy
populations with significantly different physical properties from one
another}.

\subsection{IllustrisTNG100} \label{sec:tng}
The IllustrisTNG100 simulation\footnote{\url{https://www.tng-project.org/}}
is a cosmological hydrodynamic simulation of comoving volume 
$(110.7\,\mpc)^3$, with a particle mass resolution of $7.6 \times 10^{6}\ M_\odot$ for dark matter and $1.4 \times 10^{6}\ M_\odot$ for baryonic particles~\citep{nelson2018, pillepich2018, springel2018, marinacci2018, naiman2018}. It improves on
the original Illustris simulation\footnote{\url{http://www.illustris-project.org}}~(\citealt{vogelsberger2014, genel2014};
public data release by~\citealt{nelson2015}), by including
magneto-hydrodynamics and updated treatments for galactic winds, metal
enrichment, and AGN feedback. Most notably, TNG uses a new implementation for
feedback from SMBH, where feedback energy is injected in the form of a kinetic
AGN-driven wind at low SMBH accretion rates~\citep{weinberger2018}. This new
implementation has been shown to alleviate discrepancies found between the
original Illustris and observations for $> 10^{13-14} M_\odot$ massive halos. 

\subsection{EAGLE} \label{sec:eagle} 
The Virgo Consortium's EAGLE
project\footnote{\url{http://www.eaglesim.org}}~\citep{schaye2015, crain2015,
mcalpine2016} is a publicly available suite of cosmological hydrodynamic
simulations constructed using {\sc Anarchy} (Dalla Vecchia et al. in prep.; 
see also Appendix A of \citealt{schaye2015}), a modified version of the 
{\sc GADGET-3} code~\citep{springel2005}. We use the L0100Ref simulation,
which has a comoving volume of $(100\,\mpc)^3$, and a baryonic mass resolution of $1.81\times 10^6M_{\sun}$. 
EAGLE has subgrid models for star formation, stellar mass loss, metal enrichment
and stellar feedback that stochastically inject thermal energy in the ISM as
in~\cite{dallavecchia2012}. The feedback energy from AGN is also added to
surrounding gas stochastically~\citep{booth2009}. Parameters of the stellar 
feedback and SMBH accretion are calibrated to broadly reproduce the $z=0$ 
stellar mass function and galaxy stellar size-stellar mass relation. Meanwhile, 
the AGN feedback efficiency is calibrated to match the SMBH-galaxy mass relation. 

\subsection{SIMBA} \label{sec:simba}
The {\sc Simba} simulation suite~\citep{dave2019}, the successor to {\sc
Mufasa}~\citep{dave2016, dave2017, dave2017a}, is a cosmological hydrodynamical
simulation constructed using the {\sc Gizmo} meshless finite mass hydrodynamics 
code~\citep{hopkins2015, hopkins2017}. Of the suite, we use
`m100n1024', which has a box size of $(100\,h^{-1}\,\mpc)^3$ and baryonic 
mass resolution of $1.82 \times 10^7\ M_\odot$. The simulation uses some of the same
subgrid models as {\sc Mufasa} for $\rm H_2$ based star formation, decoupled
two-phase winds for star formation driven galactic winds with velocity and mass loading 
calibrated to reproduce the FIRE simulations \citep{muratov2015, angles-alcazar2017b}, and feedback from 
Type I supernovae and AGB stars. {\sc Simba} uses a two-mode SMBH accretion 
model, torque-limited accretion for cold gas~\citep{angles-alcazar2017} and 
Bondi-based accretion for hot gas, and AGN feedback in jet, radiative, and
X-ray modes. 

\begin{figure}
\begin{center}
    \includegraphics[width=0.45\textwidth]{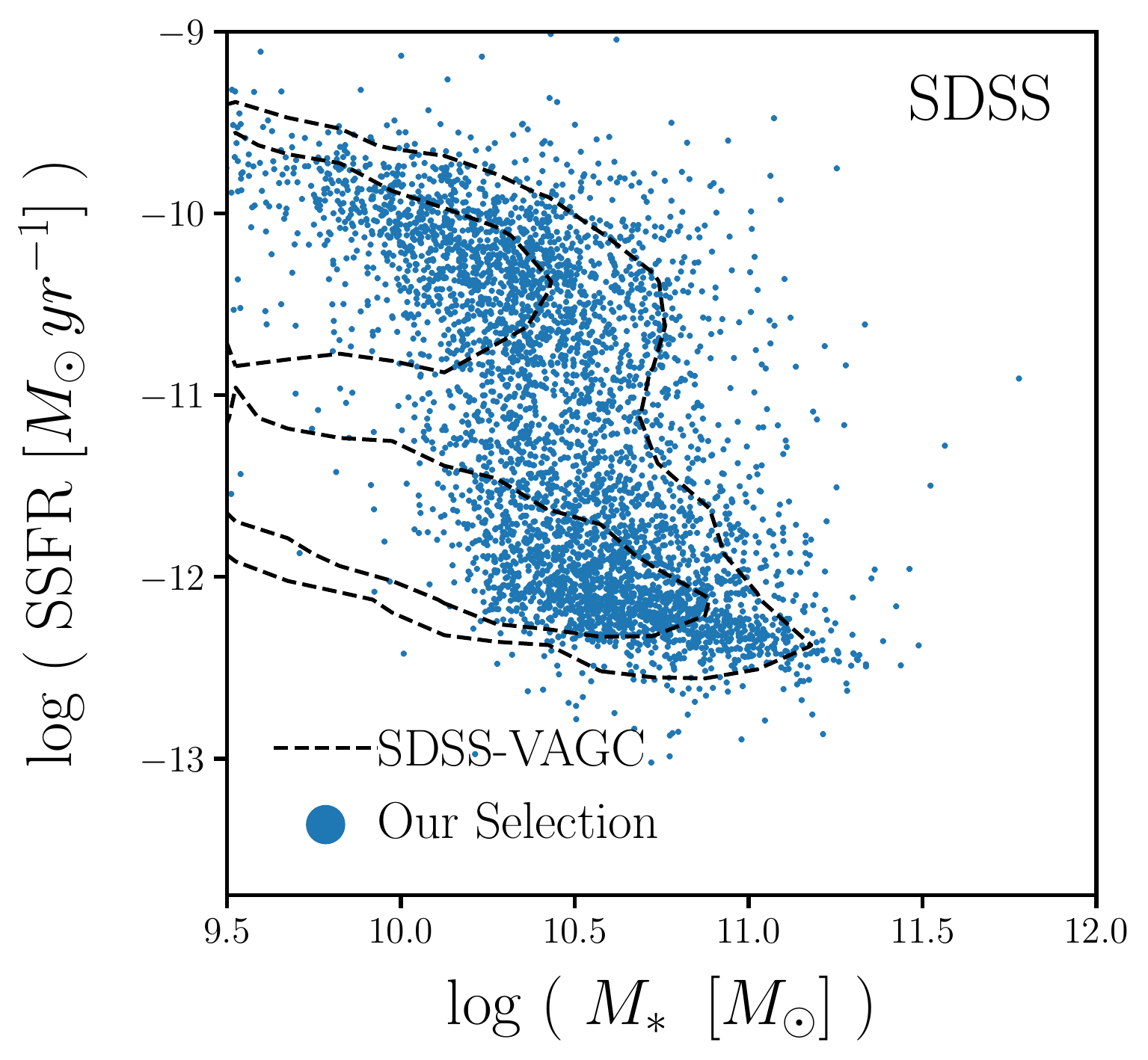} 
    \caption{\label{fig:sdss_sel}
    We derive our observational sample (blue) from the \cite{tinker2011}
    SDSS sample (black dashed) by impose $M_r < -20$, $M_{FUV}
    < -13.5$  and $M_{NUV} < -14.0$ completeness limits. 
    We describe the galaxy sample and completess limits in Section~\ref{sec:obs}. 
    $M_*$ is estimated using $\mathtt{kcorrect}$ and SSFR is from
    \cite{brinchmann2004}.
    Our SDSS sample has 4,451 galaxies that includes both
    star-forming and quiescent galaxies with $M_* \gtrsim 10^{10}M_\odot$.
    }
\end{center}
\end{figure}

\subsection{SDSS Galaxies} \label{sec:obs} 
For our observations, we 
begin with the volume-limited \cite{tinker2011} sample derived from the SDSS
DR7~\citep{abazajian2009} NYU Value-Added Galaxy
Catalog~\citep[VAGC;][]{blanton2005}, which has a $M_* > 10^{9.7} M_\odot$
completeness limit. 
However, rather than $M_*$, we focus on observables that can be consistently
defined and derived in both simulations and observations: the $r$-band absolute
magnitude, $M_r$, the optical $\gr$ color, and the $\fnuv$ color. 
We use $FUV$, $NUV$, $r$ and $g$ band absolute magnitudes from the NASA-Sloan
Atlas\footnote{\url{http://nsatlas.org/}} (NSA), which is a re-reduction of SDSS DR8
\citep{aihara2011} that includes an improved background subtraction~\citep{blanton2011} 
and near and far UV photometry from GALEX. These absolute magnitudes are
derived using $\mathtt{kcorrect}$~\citep{blanton2007a}, assuming
a~\cite{chabrier2003} initial mass function. 

We impose a $M_r < -20$ completeness limit on the \cite{tinker2011} sample as
well as completeness limits in the $FUV$ and $NUV$ bands. 
$\mathtt{kcorrect}$ UV absolute magnitudes are poorly constrained for
galaxies with low UV fluxes. 
We compare the reconstructed $FUV$ and $NUV$ fluxes from
$\mathtt{kcorrect}$ to the measured fluxes and determine the flux limits
above which the fluxes are in good agreement. 
The flux limits correspond to completeness limits of $M_{FUV} < -13.5$  and
$M_{NUV} < -14.0$. 
In Figure~\ref{fig:sdss_sel}, we present the $M_*$-SSFR relation of our
observational sample (blue). 
We include the original \cite{tinker2011} SDSS sample (black dash) for
comparison.  
In total, our SDSS sample has 4,451 star-forming and quiescent galaxies
with $M_* \gtrsim 10^{10}M_\odot$.

\subsection{Forward Modeling Observations} \label{sec:fm} 
One of the main goals of this work is to conduct an ``apples-to-apples''
comparison between the simulations and observations. 
A crucial step in this comparison is to \emph{forward model} the
observables from the simulations. 
The simulations can then be directly compared to observations in
observational-space, instead of relying on measured galaxy properties,
which are impacted by variations, inconsistencies, and biases of different
methods~\citep{dickey2020}. 
The comparison can also include selection functions and observational systematic
effects through the forward model. 
In this work, we use $r$-band luminosity ($M_r$), optical color ($\gr$),
and UV color ($\fnuv$) as our observables. 

First, we construct SEDs for all of the simulated galaxies based on their star
formation and metallicity histories (SFH and ZH) using the Flexible Stellar Population Synthesis
model~\citep[$\mathtt{FSPS}$;][]{conroy2009, conroy2010} with the MILES
spectral library~\citep{sanchez_blazquez2006}, MIST
isochrones~\citep{paxton2011, paxton2013, paxton2015, choi2016, dotter2016},
and \cite{chabrier2003} initial mass function.
For each simulated galaxy, we bin the total stellar mass formed by age ($t$) and metallicity
($Z$). We use the same $t$, $Z$ grid for all of the simulations
to account for the variable time and mass resolutions. 
We assume each $(t, Z)$ bin is a single stellar population and generate a
spectrum assuming using $\mathtt{FSPS}$ and take the mass-weighted linear
combination of them to produce the galaxy SED. 
For further details on how we construct the SEDs, we refer readers to
Starkenberg et al. (in prep.).

Next, we apply dust attenuation to the SEDs using the \eda~prescription, which 
assigns dust attenuation curves to each simulated galaxy based on its physical
properties and \eda~model parameters (Section~\ref{sec:dem}). 
We then convolve the attenuated SEDs with the transmission curves of the GALEX
$FUV$, GALEX $NUV$, SDSS $g$, and SDSS $r$ broadband filter to construct the
observables. 
We add realistic noise to $M_r$, $\gr$, and $\fnuv$ by sampling from the
observed uncertainty distributions of the NASA Sloan-Atlas.
Lastly, we apply the same $M_r < -20$, $M_{FUV} < -13.5$, and $M_{NUV} < -14$
absolute magnitude completeness limits of our SDSS sample to the simulated 
galaxies. 

\begin{figure}
\begin{center}
\includegraphics[width=0.9\textwidth]{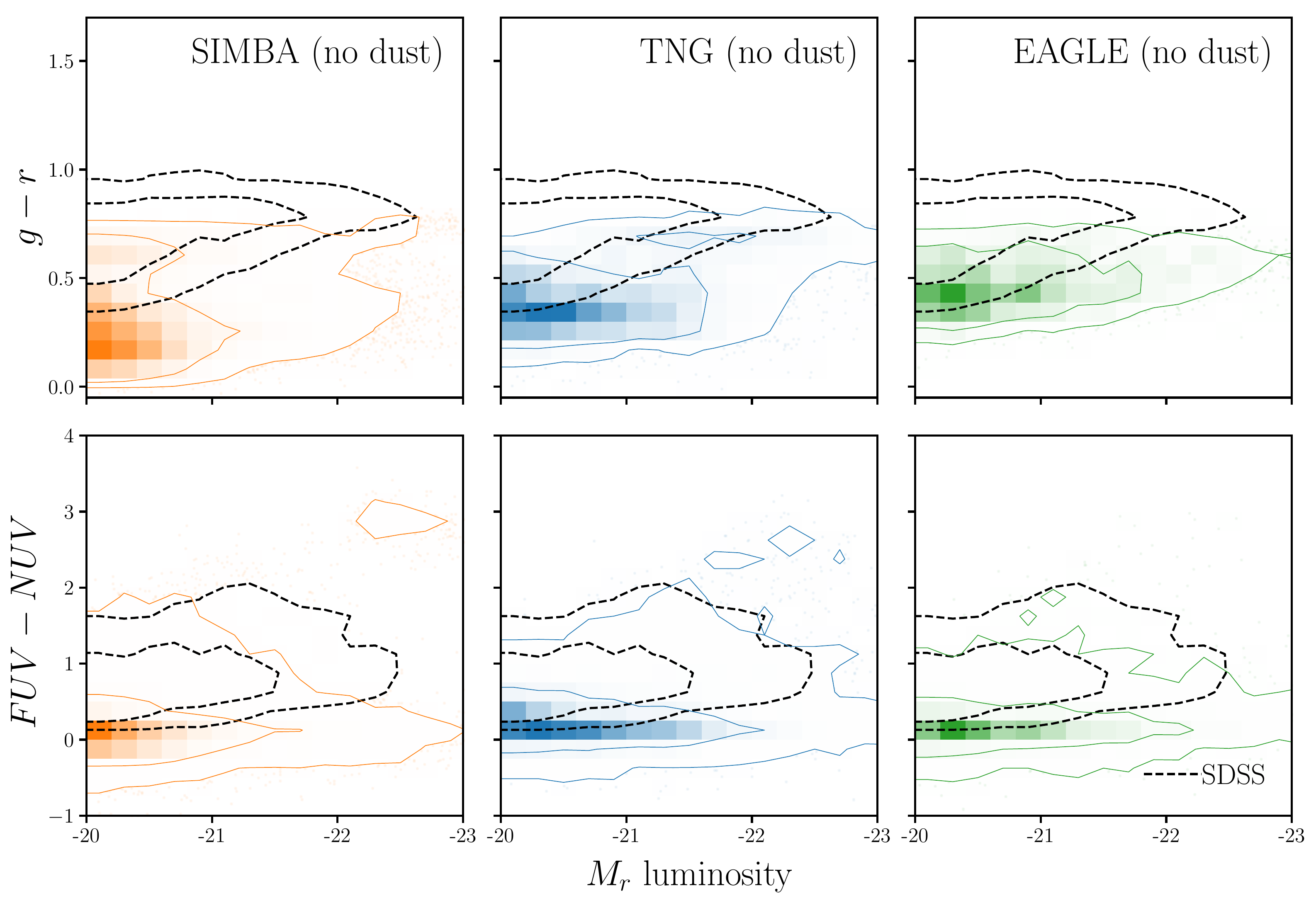} 
    \caption{\label{fig:obs}
    We present the forward modeled optical and UV color-magnitude relations
    of SIMBA (left), TNG (center), and EAGLE (right) galaxies
    \emph{assuming no dust attenuation}. We present $(\gr)-M_r$ in the top
    panels and $(\fnuv)-M_r$ in the bottom panels. The contours represent
    the 68 and 95\% of the distribution. We derive observables $M_r$, $\gr$, 
    and $\fnuv$ for the simulations using our forward model
    (Section~\ref{sec:fm}). For comparison, we include the color-magnitude
    relations of our SDSS sample (black dashed; Section~\ref{sec:obs}). 
    {\em Without dust attenuation, the hydrodynamical simulations do not
    reproduce the SDSS optical or UV color-magnitude relations.}
    }
\end{center}
\end{figure}

In Figure~\ref{fig:obs}, we present the forward modeled optical and UV
color-magnitude relations, $(\gr)-M_r$ (top) and $(\fnuv)-M_r$ (bottom),
for simulated galaxies in SIMBA (left), TNG (center) and EAGLE (right)
\emph{assuming no dust attenuation}. We mark the 68 and 95\% contours and
include, for reference, the optical and UV color-magnitude relations of our
SDSS sample (black dashed). 
The comparison to SDSS observations clearly demonstrates
that {\em without dust attenuation, the hydrodynamical simulations do not
reproduce the observed optical or UV color-magnitude relations.}

%% file: dem.tex
\section{The Empirical Dust Attenuation Framework} \label{sec:dem}
In this section, we describe the Empirical Dust Attenuation (\eda)
framework and present the \eda~prescription used in this work to apply 
dust attenuation to our simulated galaxies.
We begin by defining the dust attenuation curve, $A(\lambda)$, as 
\begin{equation} \label{eq:full_atten}
    F_o (\lambda) = F_i (\lambda) 10^{-0.4 A(\lambda)}
\end{equation}
where $F_o$ is the observed flux and $F_i$ is the intrinsic flux. We normalize
the attenuation to the $V$ band attenuation, 
\begin{equation} 
    A(\lambda) = A_V \frac{k(\lambda)}{k_V}
\end{equation}
so that $A_V$ determines the amplitude of the attenuation, while $k(\lambda)$
determines the wavelength dependence. 
The \eda~assigns a $A_V$ and $k(\lambda)$ for each simulated galaxy. 
For $A_V$, we use the slab model~\citep[\eg][]{somerville1999, somerville2012},
where $A_V$ is a 
function of galaxy inclination, $i$, and galaxy properties: 
\begin{equation} \label{eq:slab}
    A_V = -2.5 \log \left[ \frac{1 - e^{-\tau_V\,\sec i}}{\tau_V\,\sec i} \right].
\end{equation}
$\tau_V$ is the $V$-band optical depth that depends linearly on $M_*$ and
$\ssfr$: 
\begin{equation} \label{eq:tauv}
    \tau_V(M_*, \sfr) = \mtaum \log \left(\frac{M_*}{10^{10} M_\odot}\right) +
    \mtaus \log \left(\frac{\ssfr}{10^{-10}yr^{-1}}\right) + c_\tau.
\end{equation}
$\mtaum$, $\mtaus$, and $c_\tau$ represent the $M_*$ dependence, the $\ssfr$
dependence, and amplitude of $\tau_V$. Since $\tau_V$ is optical depth, we
impose a $\tau_V \ge 0$ limit.
For each galaxy, we uniformly sample $\cos i$ from 0 to 1 to introduce
stochasticity. 
This produces significant variance in $A_V$ so galaxies with the same
properties do not have identical dust attenuation.

Our $\tau_V$ parameterization is based on correlations between dust attenuation
and galaxy properties that have been established by
observations~\citep[\eg~][]{garn2010, battisti2016, salim2020}.
Previous works have parameterized dust attenuation based on other galaxy properties
such as gas density, gas metallicity, or star-gas geometry, motivated by
the fact that dust attenuation on small scales depends on local stellar and gas
properties~\citep[\eg][]{somerville1999, somerville2012, steinacker2013,
camps2015, narayanan2018, trayford2020, vogelsberger2020}. 
Galaxies in the SIMBA, TNG, and EAGLE, however, have
substantially different gas masses and metallicites~\citep[][Maller \etal~in prep.]{dave2020}.  
If we were to parameterize $\tau_V$ using these properties, their differences
would dominate any comparison of dust attenuation.
In Appendix~\ref{sec:slab}, we confirm the correlation between $A_V$ and
the properties $M_*$ and $\ssfr$ in the \cite{salim2018} GSWLC2 sample
(Figure~\ref{fig:dep}). 

In our \eda, we use the slab model because it provides a simple
prescription for generating a distribution of $A_V$ that depends on
randomly sampled $i$, with loose physical motivations.
For star-forming galaxies, which typically have disc-like morphologies, the
slab model produces $A_V$ that is correlated with $i$ in a way consistent
with observations: edge-on galaxies have higher $A_V$ than face-on
galaxies~\citep[\eg][]{conroy2010, wild2011, battisti2017, salim2020}.
Nevertheless, the slab model is a simplification. 
In reality, $A_V$ depends on the detailed star-to-dust geometry.
Furthermore, we assign $A_V$ to all galaxies, not just star-forming.
For quiescent galaxies, which typically have elliptical morphologies, the
slab model serves only as an \emph{empirical} prescription for statistically 
sampling $A_V$. 
The~\eda~seeks to assign an accurate distribution of dust
attenuation curves for an ensemble of galaxies --- \emph{not} to accurately
model dust attenuation for individual galaxies.
In this regard, we demonstrate in Appendix~\ref{sec:slab} that the slab model
can match the observed distribution of $A_V$, even for samples that
include quiescent galaxies.

For the wavelength dependence of the attenuation curve, $k(\lambda)$, we
use \cite{noll2009} parameterization: 
\begin{equation} \label{eq:noll}
    k(\lambda) = \left(k_{\rm Cal}(\lambda) + D(\lambda)\right) \left(
    \frac{\lambda}{\lambda_V} \right)^\delta.
\end{equation}
Here $k_{\rm Cal}(\lambda)$ is the \cite{calzetti2001} curve: 
\[
    k_{\rm Cal}(\lambda) = 
    \begin{cases} 
        2.659 (-1.857 + 1.040/\lambda) + R_V, & 6300 A \le \lambda \le
        22000 A \\ 
        2.659 (-2.156 + 1.509/\lambda - 0.198/\lambda^2 + 0.011/\lambda^3) +
        R_V & 1200 A \le \lambda \le 6300 A
    \end{cases}
\]
where $\lambda_V = 5500 A$ is the $V$ band wavelength and $\delta$ is the slope
offset of the attenuation curve from $k_{\rm Cal}$. Since $\delta$ correlates 
with galaxy properties~\citep[\eg][see also Appendix~\ref{sec:slab}]{wild2011, battisti2016, leja2017, salim2018},
we parameterize $\delta$ with a similar $M_*$ and $\ssfr$ dependence as
$\tau_V$:  
\begin{align} \label{eq:delta}
    \delta(M_*, \sfr) &= \mdeltam \log \left(\frac{M_*}{10^{10}
    M_\odot}\right) + \mdeltas \log \left(\frac{\ssfr}{10^{-10}yr^{-1}}\right)
    + c_\delta.
\end{align}
$D(\lambda)$ in Eq.~\ref{eq:noll} is the UV dust bump, which we parameterize using
the standard Lorentzian-like Drude profile:
\begin{equation}
    D(\lambda) = \frac{E_b(\lambda~\Delta \lambda)^2}{(\lambda^2 -
    \lambda_0^2)^2 + (\lambda~\Delta \lambda)^2}
\end{equation}
where $\lambda_0 = 2175 \AA$, $\Delta \lambda = 350\AA$, and $E_b$ are the
central wavelength, full width at half maximum, and strength of the bump,
respectively. 
We include the UV dust bump since we use UV color as one of our observables.
\cite{kriek2013} and \cite{tress2018} find that $E_b$ correlates with $\delta$ for star-forming galaxies at $z{\sim}2$.
\cite{narayanan2018} confirmed this dependence in simulations. 
Hence, we assume a fixed relation between $E_B$ and $\delta$: $E_b =
-1.9~\delta + 0.85$~\citep{kriek2013}. 
Allowing the slope and amplitude
of the $E_B$ and $\delta$ relation to vary does {\em not} impact our results;
however, we also do not derive any meaningful constraints on them. In
Table~\ref{tab:free_param}, we list and describe all of the free parameters of
our \eda~prescription. 


$\ssfr$ of galaxies are used to calculate $\tau_V$ and $\delta$ in
Eqs.~\ref{eq:tauv} and~\ref{eq:delta}. However, due to mass and temporal resolution limits,
some galaxies in the simulations have $\sfr=0$ --- \ie~an unmeasurably low
SFR~\citep{hahn2019c}. They account for 17, 19, 9\% of galaxies
in SIMBA, TNG, and EAGLE, respectively. Since Eqs.~\ref{eq:tauv}
and~\ref{eq:delta} depend on $\log\ssfr$, they cannot be used in the equations
to derive $\tau_V$ and $\delta$ for these galaxies. To account for this issue,
we assign $\sfr_{\rm min}$, the minimum non-zero $\sfr$ in each simulation, to
$\sfr=0$ galaxies when calculating $\tau_V$ and $\delta$. For SIMBA, TNG, and
EAGLE, $\sfr_{\rm min}=0.000816$, $0.000268$, and $0.000707 M_\odot/yr$,
respectively. Although 
this assumes that $\sfr=0$ galaxies have similar dust properties as the galaxies 
with $\sfr = \sfr_{\rm min}$, since the simulations have very low $\sfr_{\rm min}$ 
we expect galaxies with $\sfr = \sfr_{\rm min}$ to have little recent
star-formation and low gas mass, similar to $\sfr=0$ galaxies. 

\begin{figure}
\begin{center}
    \includegraphics[width=0.6\textwidth]{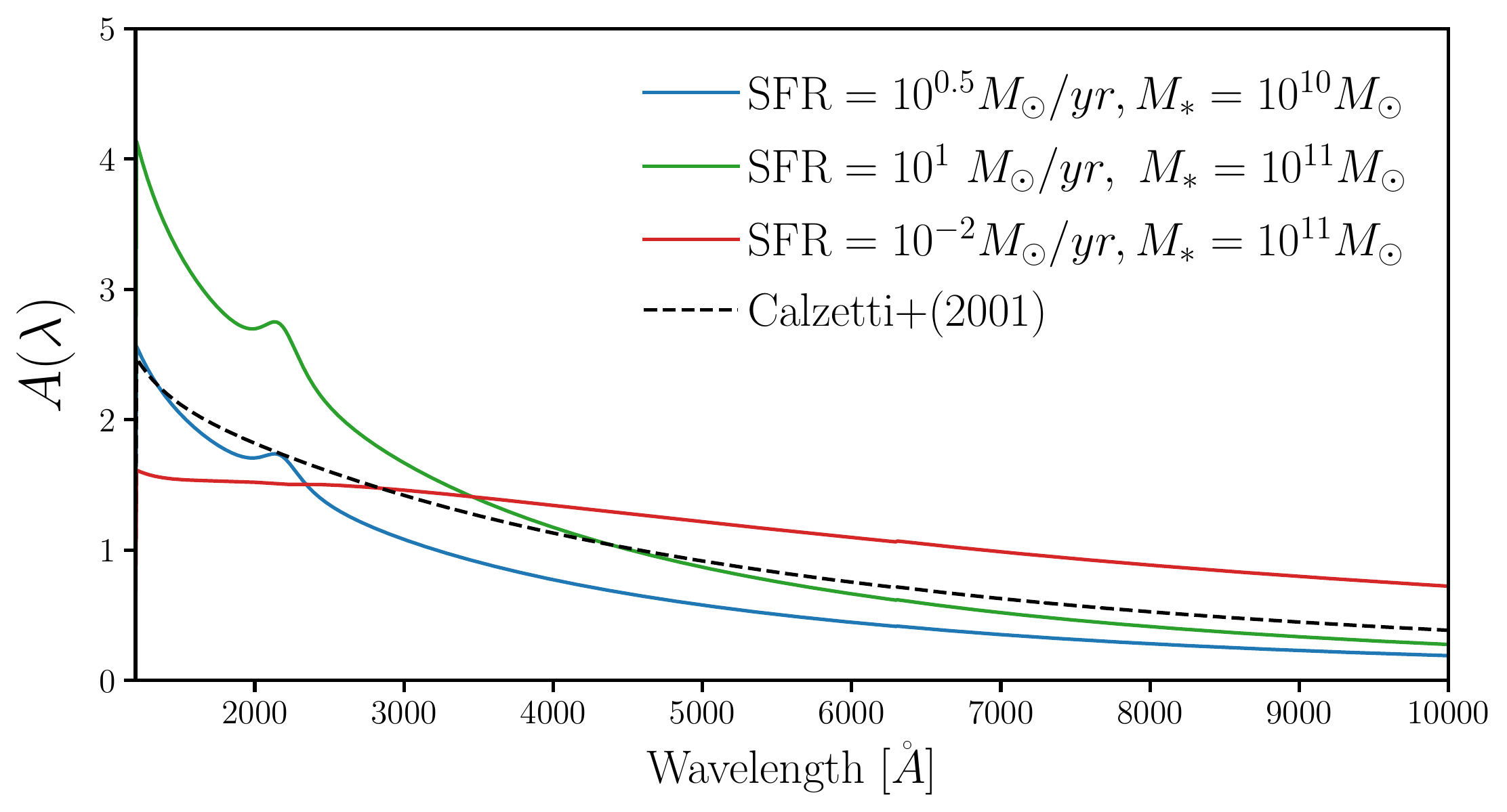}
    \caption{\label{fig:dem_av}
    Attenuation curves, $A(\lambda)$, assigned by our Empirical Dust
    Attenuation (\eda) prescription to edge-on galaxies with different $\sfr$ and
    $M_*$ values for an arbitrary set of \eda~parameters. We include
    $A(\lambda)$ for star-forming galaxies with $\{M_*, \sfr\} =
    \{10^{10}M_\odot, 10^{0.5}M_\odot/yr\}$ (blue), $\{10^{11}M_\odot, 10^{1}
    M_\odot/yr\}$ (green) and a quiescent galaxy with $\{10^{11}M_\odot,
    10^{-2}M_\odot/yr\}$ (red). We set $i=0$ for
    all the galaxies in the figure for simplicity but in practice the
    \eda~uniformly samples $\cos i$ from 0 to 1 for each galaxy.
    For comparison, we include the \cite{calzetti2001} attenuation curve.
    {\em The \eda~provides a flexible prescription for assigning dust
    attenuation to galaxies based on their physical properties ($M_*$ and
    $\ssfr$) and the \eda~parameters.}
    } 
\end{center}
\end{figure}

In summary, to apply the \eda~to a simulated galaxy population, we first
assign a randomly sampled $i$ to each galaxy ($\cos i$ uniformly sampled from 0 to 1).
$\tau_V$ and $\delta$ are calculated for
the galaxy based on its $M_*$,
$\ssfr$ and the \eda~parameters. 
We then calculate $A_V$ and $k(\lambda)$ to determine $A(\lambda)$ for each galaxy.
Afterwards, we attenuate the galaxy SEDs using Eq.~\ref{eq:full_atten} and use
the attenuated SEDs to calculate the observables: $g, r, NUV$, and $FUV$
absolute magnitudes. 
In Figure~\ref{fig:dem_av}, we present attenuation curves, $A(\lambda)$,
generated by the \eda~for galaxies with different $\sfr$ and $M_*$ values.  
We present star-forming galaxies with $\{M_*, \sfr\} = \{10^{10}M_\odot,
10^{0.5}M_\odot/yr\}$ (blue), $\{10^{11}M_\odot, 10^{1} M_\odot/yr\}$
(green) and a quiescent galaxy with $\{10^{11}M_\odot, 10^{-2}M_\odot/yr\}$
(red).
We use an arbitrary set of \eda~parameters ($\mtaum, \mtaus, c_\tau,
\mdeltam, \mdeltas, c_\delta$) within the prior range listed in
Table~\ref{tab:free_param}. 
We set $i=0$ (edge-on) for all $A(\lambda)$ in Figure~\ref{fig:dem_av} for
simplicity.
For comparison, we include the \cite{calzetti2001} attenuation curve. 
The \eda~produces attenuation curves with a wide range of amplitudes and slopes
for galaxies based on their physical properties. 

\begin{table}
    \caption{Free parameters of the Empirical Dust Attenuation Model}
    \begin{center}
        \begin{tabular}{ccc} \toprule
            Parameter & Definition & prior\\[3pt] \hline\hline
            $\mtaum$ & $M_*$ dependence of the optical depth, $\tau_V$ & flat $[-5., 5.]$\\
            $\mtaus$ & $\ssfr$ dependence of $\tau_V$  & flat $[-5., 5.]$\\
            $c_{\tau}$ & amplitude of $\tau_V$ & flat $[0., 6.]$\\
            $\mdeltam$ & $M_*$ dependence of $\delta$, the attenuation curve slope offset & flat $[-4., 4.]$\\
            $\mdeltas$ & $\ssfr$ dependence of $\delta$ & flat $[-4., 4.]$\\
            $c_{\delta}$ & amplitude of $\delta$ & flat $[-4., 4.]$\\
            \hline
        \end{tabular} \label{tab:free_param}
    \end{center}
\end{table}

%% file: abc.tex
\section{Likelihood-Free Inference: Approximate Bayesian Computation} \label{sec:abc}
With our forward model, which includes the \eda~prescription for dust
attenuation, we can now generate synthetic observations for simulated
galaxies and make an ``apples-to-apples'' comparison to SDSS. Next, we want
to use this comparison to infer the posterior probability distribution of
the \eda~parameters. Typically in astronomy, this inference is done
assuming a Gaussian likelihood to compare the ``summary statistic''
(\eg~SMF) of the model to observations and some sampling method (\eg~Markov
Chain Monte Carlo) to estimate the posterior distribution. The functional form of the
likelihood, however, depends on the summary statistic and assuming an
incorrect form of the likelihood can significantly bias the inferred
posteriors~\citep[\eg][]{hahn2019}. In this work, we use the optical and UV
color-magnitude relations as our summary statistic. Since this statistic is
a three-dimensional histogram, the likelihood is {\em not} Gaussian.
Furthermore, since the bins are not independent, the true likelihood is
difficult to analytically write down.

Rather than \emph{incorrectly} assuming a Gaussian likelihood or attempting
to estimate the true likelihood of the optical and UV color-magnitude
relations, we use Approximate Bayesian Computation~\citep[hereafter
ABC;][]{diggle1984, tavare1997, pritchard1999, beaumont2009, delmoral2012}
for our inference. 
ABC is a likelihood-free (or ``simulation-based'') parameter inference
framework that approximates the posterior probability distribution, $p(\theta\given{\rm data})$, without
requiring evaluations of the likelihood.  Instead, ABC only requires a forward
model of the observed data, a prior that can be sampled, and a distance metric
that quantifies the ``closeness'' to the observed data. 
Since ABC does not require evaluating the likelihood, it does not assume
any functional form of the likelihood so we avoid any biases from such
assumptions. 
It also expands the summary statistics we can use to infer the posteriors and,
therefore, provides a general inference framework for a forward modeling 
approach. 

In the simplest version of ABC, with rejection sampling~\citep{pritchard1999}, a proposal set of parameter values are drawn
from the prior. The forward model is run with the proposal parameter values.
The output of the forward model is then compared to the observed data using
a distance metric. 
If the distance is within some small threshold, we keep the proposed
parameters; otherwise, we discard them.  Proposals are
drawn until enough pass the threshold to sample the posterior. A
rejection sampling framework requires a large number of evaluations of the
forward model, which
can be computationally costly. Many variations of ABC with more efficient
sampling strategies have now been applied to astronomy and
cosmology~\citep[\eg][]{cameron2012, weyant2013, ishida2015, lin2016, alsing2018}.
Among these methods, we use ABC with Population Monte Carlo (PMC) 
importance sampling~\citep{hahn2017a, hahn2017b, hahn2019a}.

ABC-PMC begins with an arbitrarily large threshold $\epsilon_1$ and $N$ proposals 
$\bar{\theta}_1$ sampled from the prior distribution. Each proposal is
assigned a weight $w^i_1 = 1/N$. Then for subsequent iterations ($n > 1$), the 
threshold, $\epsilon_n$, is set to the median distance of the previous iteration's
proposals. New proposals are drawn from the previous iteration's proposals perturbed 
by a kernel and kept if their distance is below $\epsilon_n$. This is repeated
until we assemble a new set of $N$ proposals $\bar{\theta}_n$. The entire
process is repeated for the next iteration until convergence is confirmed. 
We use the Python implementation of
\cite{akeret2015}\footnote{https://abcpmc.readthedocs.io/en/latest/index.html}.
For further details on the ABC-PMC implementation, we refer readers to \cite{hahn2017b}
and \cite{hahn2019a}.

\begin{figure}
\begin{center}
    \includegraphics[width=0.9\textwidth]{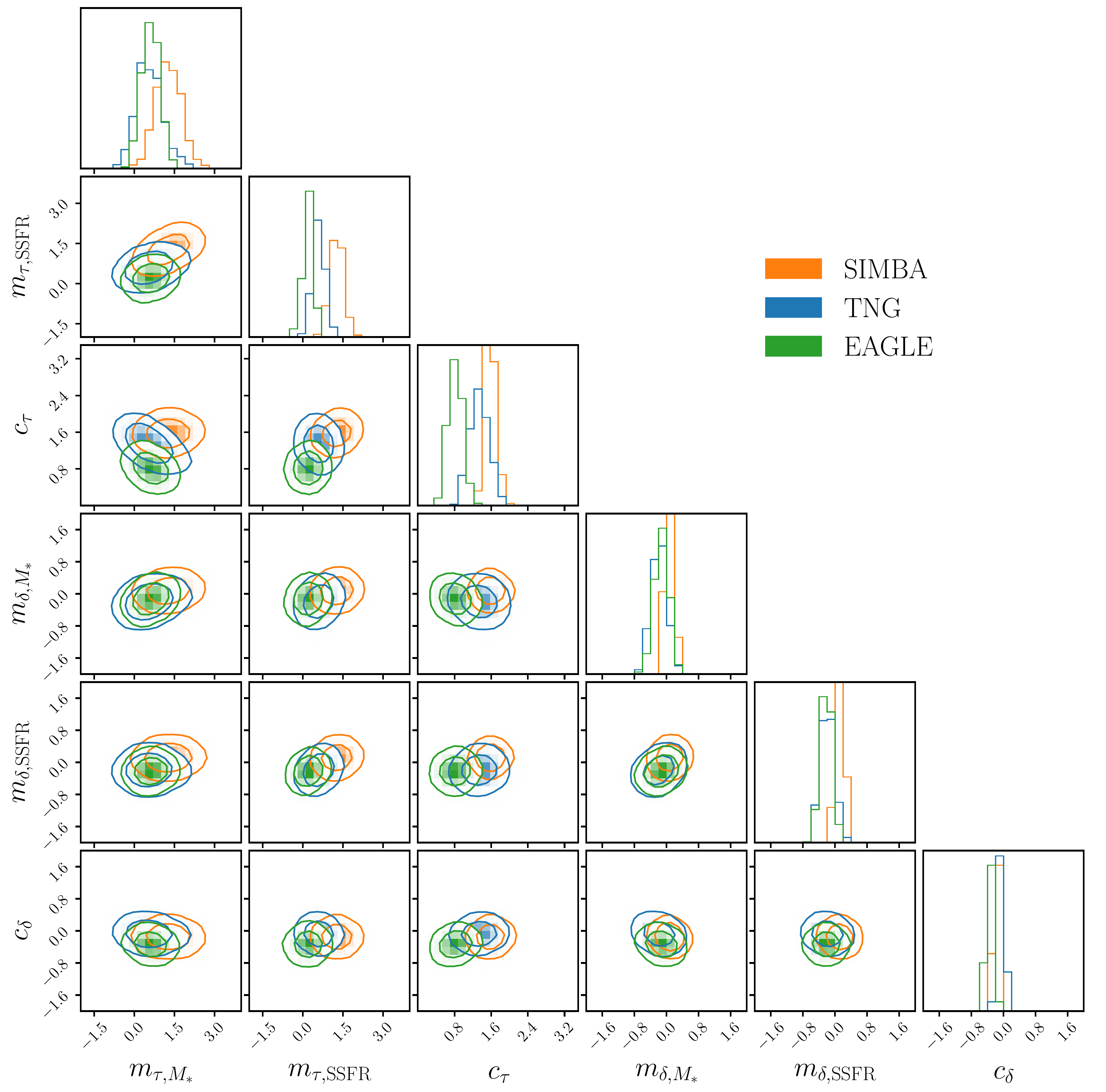}
    \caption{\label{fig:abc}
    Posterior distributions of the \eda~parameters for the SIMBA (orange), TNG
    (blue), and EAGLE (green) hydrodynamical simulations derived from comparing
    the simulations to SDSS with a forward modeling approach. 
    The \eda~parameters determine the $M_*$ dependence, $\ssfr$ dependence, and
    amplitude of $\tau_V$ and $\delta$ (Table~\ref{tab:free_param}). 
    The contours mark the $68$ and $95$ percentiles of the distributions. 
    The posteriors are derived from likelihood-free inference using Approximate
    Bayesian Computation with Population Monte Carlo (Section~\ref{sec:abc}). 
    }
\end{center}
\end{figure}

In this work, we use ABC-PMC with uninformative uniform priors on each of
the \eda~parameters and choose ranges that encompass constraints in the
literature.
The prior ranges of $\mtaum, \mtaus, c_\tau$ include the $A_V$ range and $M_*$
and $\sfr$ dependence of \cite{narayanan2018} and \cite{salim2020}. 
Meanwhile, the prior ranges of $\mdeltam, \mdeltas, c_\delta$ include the
$\delta$ range and $M_*$ and $\sfr$ dependence of \cite{leja2017} and
\cite{salim2018}. 
We list the range of the priors in Table~\ref{tab:free_param}. 
We use the forward model described in Section~\ref{sec:fm}, where we construct
SEDs for every simulated galaxy from SIMBA, TNG, and EAGLE, apply dust
attenuation with our \eda, calculate the observables ($M_r$, $\gr$, and
$\fnuv$), add realistic noise, and apply a $M_r < -20$ completeness limit. 
We use the optical and UV color-magnitude relation, $(\gr)- M_r$ and
$(\fnuv)-M_r$ as our summary statistic to fully exploit the $(M_r, \gr,
\fnuv)$ observational-space. We measure the color-magnitude relations by
calculating the number density in bins of $(\gr, \fnuv, M_r)$ with widths
$(0.0625, 0.25, 0.5)~mags$. For our distance metric, $\rho$, we use the L2
norm between the number density of the SDSS observation, $n^{\rm SDSS}$ and
of our forward model, $n^{\rm FM}(\theta_{\rm \eda})$: 
\begin{equation} \label{eq:distance}
    \rho(\theta_{\rm \eda}) = \sum\limits_{i,j} \left[n_{ij}^{\rm SDSS} -
    n_{ij}^{\rm FM}(\theta_{\rm \eda}) \right]^2.
\end{equation}
In Figure~\ref{fig:abc}, we present the posterior distributions of the \eda~parameters
derived using ABC-PMC for the SIMBA (orange), TNG (blue), and EAGLE (green).
The contours mark the $68$ and $95$ percentiles of the distributions. 

%% file: results.tex
\begin{figure}
\begin{center}
    \includegraphics[width=0.9\textwidth]{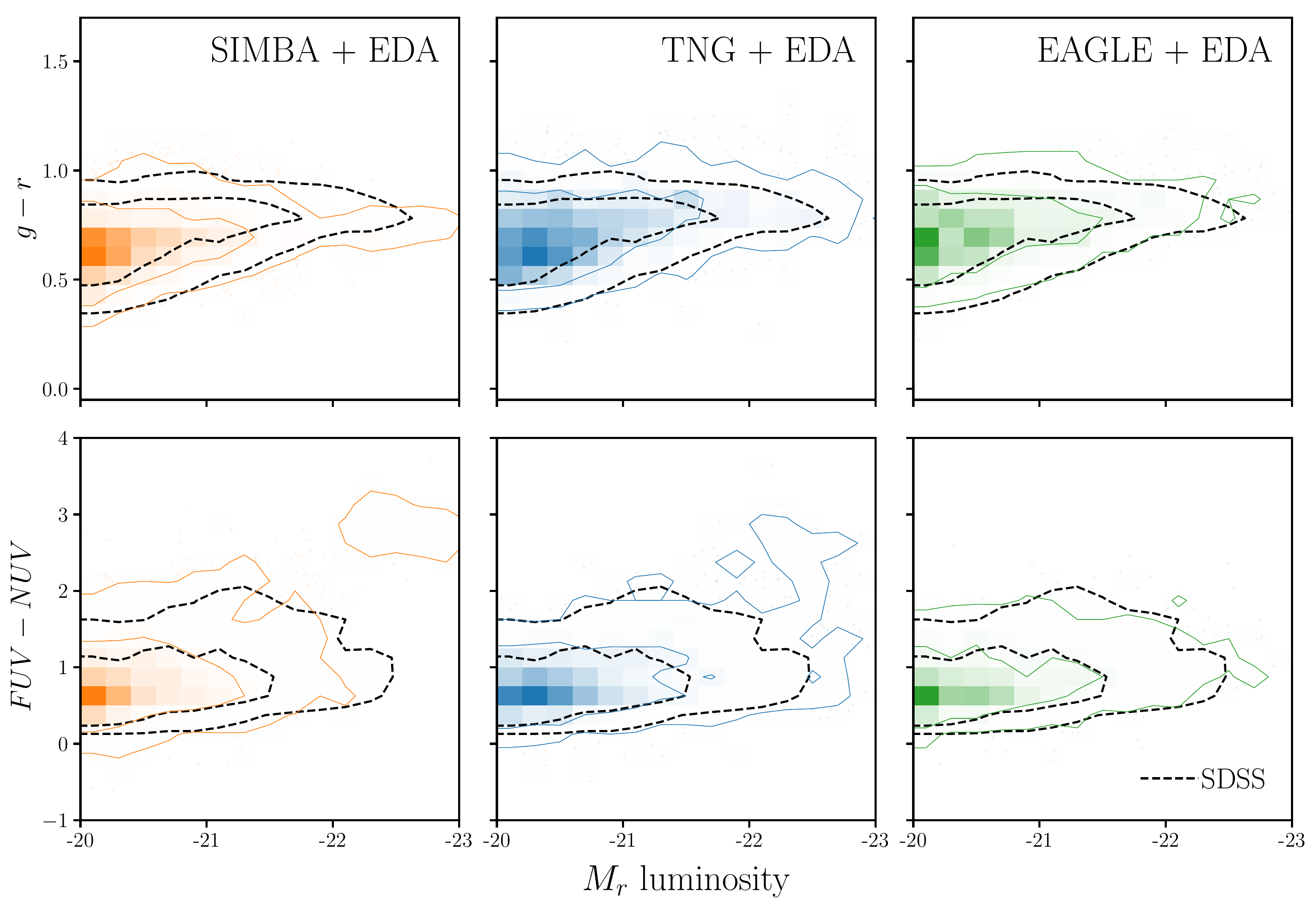}
    \caption{\label{fig:dem}
    The optical, $(\gr) - M_r$ (top), and UV, $(\fnuv) - M_r$ (bottom),
    color-magnitude relations predicted by our \eda~prescription
    for the SIMBA (orange), TNG (blue), and EAGLE (green) hydrodynamical
    simulations. For the \eda~parameters of each simulation, we use the
    median of the posterior distributions inferred using ABC. For
    comparison, we include the color-magnitude relations of SDSS (black
    dashed). Comparing the color-magnitude relations above to those without
    dust attenuation in Figure~\ref{fig:obs}, we see that dust
    \emph{dramatically} impacts the color-magnitude relations. 
    Dust attenuation must be accounted for when interpreting and comparing
    simulations. Furthermore, with our \eda~prescription, all three
    simulations reproduce the color-magnitude relations of SDSS
    observations.  \emph{Since the different simulations can reproduce
    observations just by varying dust, dust significantly limits our ability to
    constrain the underlying physical processes of galaxy formation
    models.}
    }
\end{center}
\end{figure}

\section{Results} \label{sec:results}
Without dust attenuation, all of the hydrodynamical simulations struggle to 
reproduce the $(\gr) - M_r$ and $(\fnuv) - M_r$ relations of SDSS (Figure~\ref{fig:obs}). 
Both in the optical and UV, the simulations predict galaxies significantly bluer
than SDSS galaxies.
The simulations also predict optically blue luminous galaxies with $M_r <
-21.5$ that are not found in the observations; this is particularly
noticeable for SIMBA and TNG. 
Simulated galaxies in SIMBA also have a significantly broader distribution
of $\gr$ colors than SDSS galaxies.
Meanwhile, all of the simulations predict a broader distribution of $\fnuv$  
color than SDSS.
In fact, SIMBA and TNG predict a significant number of luminous galaxies,
$M_r < -22$, with $\fnuv > 2$ colors, beyond SDSS observations.

\emph{With our \eda~prescription, all three simulations produce
color-magnitude relations that are highly consistent with SDSS observations.}
In Figure~\ref{fig:dem}, we present the optical and UV color-magnitude
relations predicted by the 
\eda~for the SIMBA (orange), TNG (blue), and EAGLE (green) simulations. 
For the \eda~parameters, we use the median values of the inferred posterior distributions (Figure~\ref{fig:abc}). 
We include the color-magnitude
relations of SDSS observations (black-dashed) for comparison. 
The contours mark the $68$ and $95$ percentiles of the distributions. 

Dust dramatically impacts the observables of simulations. 
The \eda~affects the optical and UV color-magnitude relations in three
major ways to produce good agreement with SDSS. 
First, the \eda~significantly reddens the simulated galaxies in the optical: 
$\gr$ colors are ${\gtrsim}0.25~mag$ redder than the optical
color-magnitude relation in Figure~\ref{fig:obs} and match the $\gr$
distribution of SDSS. 
Second, the \eda~reddens non-quiescent ($\log\ssfr > -11$) galaxies in the UV
by ${\gtrsim}0.5~mag$. 
While quiescent galaxies have intrinsically red UV colors that generally agree
with SDSS ($\fnuv > 0.5$), the rest of the galaxies are intrinsically
bluer in the UV than observations. 
Lastly, the \eda~attenuates non-quiescent galaxies so there are no longer
luminous galaxies that are blue in the optical or UV --- consistent with observations.

\begin{figure}
\begin{center}
    \includegraphics[width=0.6\textwidth]{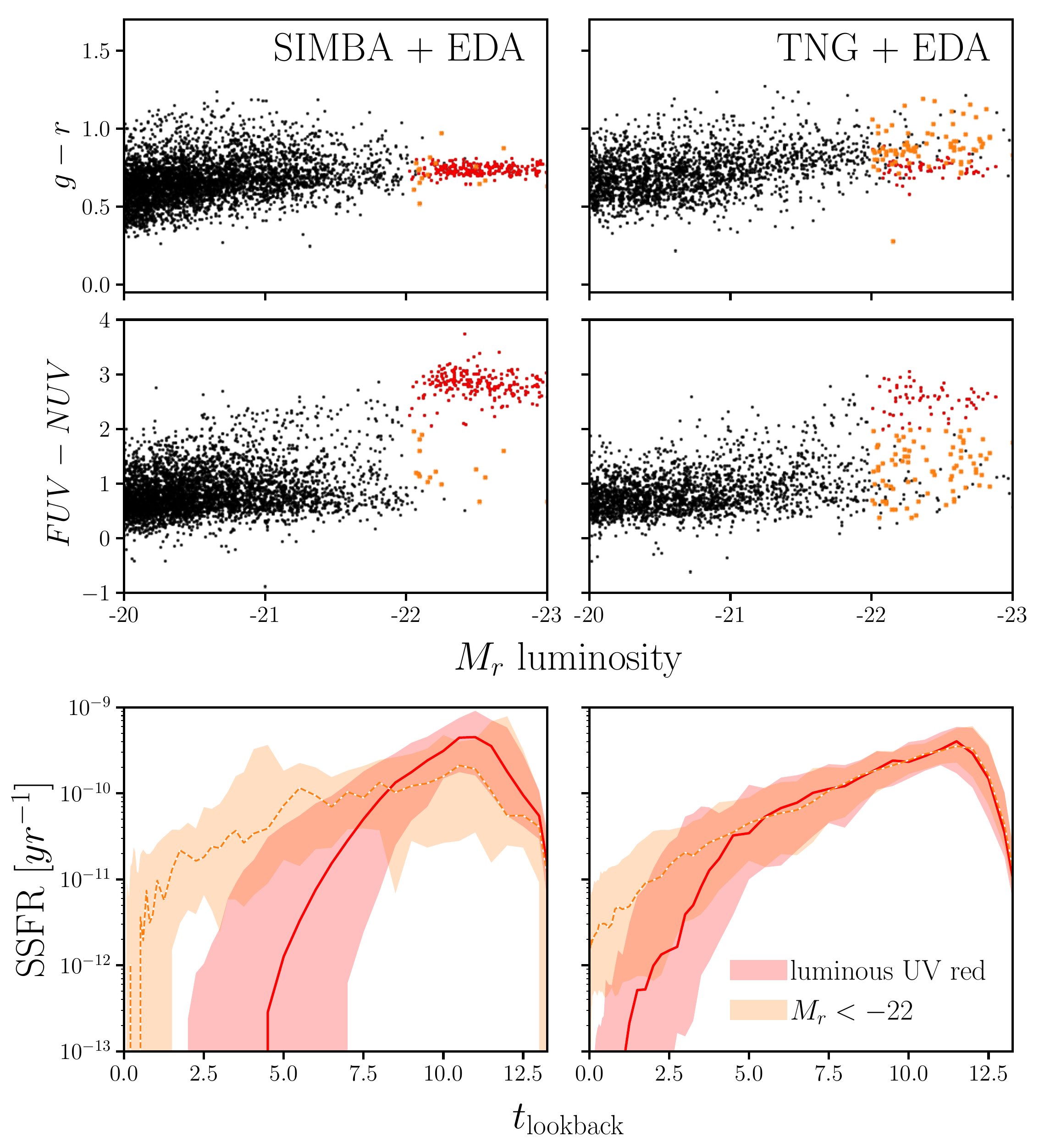}
    \caption{\label{fig:uv_sfh}
    The SFHs of luminous UV-red galaxies (red) in SIMBA (left) and TNG
    (right) that cause the discrepancy between the color-magnitude
    relations predicted by the \eda~and SDSS observations. 
    We include the SFHs of quiescent galaxies with matching luminosities, 
    $M_r < -22$, for comparison (orange). 
    The top and center panels mark the luminous UV-red and the other
    quiescent $M_r < -22$ galaxies in the \eda~predicted optical and UV
    color-magnitude relations, respectively. 
    In the bottom panels, we present the median SSFH of these galaxies 
    with the shaded regions representing the 68 percentiles of the SSFH. 
    In both TNG and SIMBA, the luminous UV red galaxies have negligible
    star formation within the last 2 Gyrs, unlike the other quiescent
    galaxies. 
    \emph{This suggests that SIMBA and TNG may be predicting luminous
    UV-red galaxies not found in observations because their
    prescription for star formation quenching is too efficient in the
    most massive galaxies.}
    }
\end{center}
\end{figure}

Despite the substantial improvement in the color-magnitude relation
agreement with the \eda, there is still one significant discrepancy: the
presence of luminous $M_r > -22$ galaxies that are UV-red, not found in
observations ($\fnuv > 2$). 
This galaxy population consists of quiescent galaxies with
$\sfr\lesssim 10^{-2}M_\odot/yr$ and is especially pronounced in the UV
color-magnitude of SIMBA but also found in TNG. 
They are also present in the UV color-magnitude predictions without dust
attenuation~(Figure~\ref{fig:obs}). 
Since they are the most luminous galaxies in our sample, dust attenuation
cannot remove them from our sample. 
Furthermore, dust reddening would only increase the discrepancies with
observations since they have intrinsic $\fnuv > 2$.
In other words, the excess luminous UV-red galaxies predicted by SIMBA and
TNG are irreconcilable with dust attenuation. 

In order to understand the origin of the luminous UV-red galaxies in SIMBA
(left) and TNG (right), we examine their star formation histories in
Figure~\ref{fig:uv_sfh}.
The top and center panels mark the luminous UV-red galaxies on the optical
and UV color-magnitude relations predicted by the \eda~(red). 
The bottom panels present the median specific SFH (SSFH), $\ssfr$ as a
function of lookback time, $t_{\rm lookback}$, with the shaded regions
representing the 68 percentile.
For comparison, we include the SSFHs of other quiescent galaxies with matching
luminosities, $\ssfr < 10^{-11}yr^{-1}$ and $M_r < -22$ (orange).
The SSFHs reveal that, unlike other quiescent galaxies, the luminous UV-red
galaxies of SIMBA and TNG have  almost no star formation within the last $t_{\rm lookback} \lesssim 2$ Gyr.
With no recent star formation contributing to the SED in $FUV$ wavelengths,
these galaxies have red $\fnuv$ color.  
These luminous UV-red galaxies in SIMBA and TNG suggest that star formation
quenching is too efficient in the most massive queiscent galaxies.

The SSFHs in Figure~\ref{fig:uv_sfh} also reveal that luminous UV-red
galaxies in SIMBA have a substantially different SSFH than other quiescent
galaxies. 
In addition to the lack of recent star formation, the luminous UV-red galaxies
also have significantly shorter star-forming timescales. 
They peak their star formation earlier than other quiescent galaxies, at
$t_{\rm lookback}\sim 11$ Gyr, and have a more rapid decline in star
formation. 
In contrast, the luminous UV-red galaxies in TNG have overall similar
SSFHs to other quiescent galaxies. 
This difference in SFH suggests that a distinct star formation quenching
mechanism is responsible for the luminous red galaxies in SIMBA. 
In another paper of the IQ series (Choi et al. in prep), we examine this
SFH difference in further detail and present its impact on the quiescent 
fraction evolution over $0 < z < 3$. 

In principle, the simulations could remove the luminous UV-red galaxies
from the sample if star formation quenching in these galaxies was {\em even}
more efficient. 
With absolutley no recent star formation, their UV SED would fall below our
$M_{FUV}$ or $M_{NUV}$ completeness limits and would not be included in the comparison.  
However, a large portion of the luminous quiescent galaxies are UV-red, so the
simulations, especially SIMBA, would struggle to reproduce the luminous end of
the SDSS red sequence and the luminosity function of the observations. 
Furthermore, reducing the total number of massive quiescent galaxies that pass
the selection function would conflict with our observations, which include a
significant number of massive quiescent galaxies (Figure~\ref{fig:sdss_sel}). 

Previous works in the literature have also compared simulations with different
dust prescriptions to observations in color-magnitude space. For EAGLE, 
\cite{trayford2015} calculated colors and luminosities with the {\sc Galaxev}
population synthesis models and a two-component screen model for dust. More
recently, \cite{trayford2017} calculated optical colors for EAGLE using {\sc
Skirt}, a Monte Carlo radiative transfer code~\citep{camps2015}, to model the
dust. At stellar masses and luminosities comparable to our SDSS sample, both 
\cite{trayford2015} and \cite{trayford2017} produce red sequences bluer than 
in GAMA observations. Also, \cite{trayford2015} predict an excess of luminous 
blue galaxies. Although a detailed comparison is difficult since both works 
compare to different observations, we note that with the \eda, EAGLE is able 
to successfully reproduce the position of the SDSS red sequence and does not 
predict a significant excess of luminous blue galaxies. Also using EAGLE and 
{\sc Skirt}, \cite{baes2019} find that they overestimate the observed cosmic 
SED (CSED) in the UV regime and produce significantly higher $\fnuv$ color 
than GAMA. The \eda~for EAGLE predicts $\fnuv$ in good agreement with SDSS. 
For TNG, \cite{nelson2018} calculate optical colors using a dust model that
includes attenuation due to dense gas birth clouds surrounding young stellar
populations and also due to simulated distribution of neutral gas and metals.
They find bluer red sequence peaks and a narrower blue cloud compared to SDSS.
We find neither of these discrepancies for the TNG+\eda. The \eda~provides a
simpler empirical framework for applying dust attenuation than the dust models
in these works. 
Yet, with its flexibility and low computation cost, we are able to fully
explore our dust parameters and produce optical and UV color-magnitude
relations that are in good agreement with observations.

\input{reproduce.tex}

\input{quiescent.tex}

\input{galaxydust.tex}

\input{discuss.tex}

%% file: reproduce.tex
\begin{figure}
\begin{center}
    \includegraphics[width=0.9\textwidth]{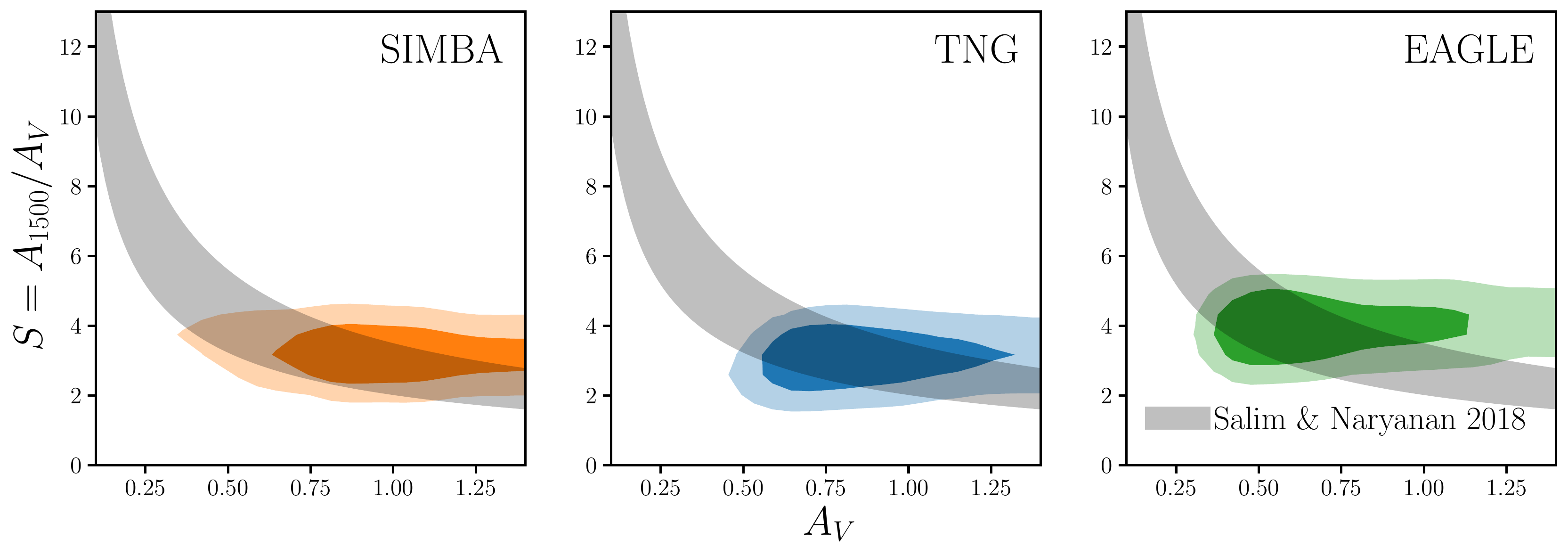}
    \caption{\label{fig:slope}
    The attenuation-slope relation of star-forming galaxies ($\ssfr >
    10^{-11}yr^{-1}$), using the attenuation curves predicted by our  
    \eda~prescription for the median posterior parameter values of SIMBA
    (left), TNG (center) and EAGLE (right). 
    For comparison, we include the observed attenuation-slope relation
    from GSWLC2~\citep{salim2020}. 
    We use $A_V$ and $S = A(1500\AA)/A_V$ as measurements of attenuation and
    slope, respectively. 
    \emph{The \eda~does not predict $A_V < 0.3$ because star-forming galaxies in
    the simulations are intrinsically too lumnious and require significant attenuation to
    match observations.}
    Beyond $A_V > 0.3$, however, there is good agreement between the
    attenuation-slope relation predicted by the \eda~and observations. 
    }
\end{center}
\end{figure}

\subsection{Comparison to Dust Observations} \label{sec:reproduce}
In addition to  reproducing observations, the \eda~assigns dust attenuation
curves to each simulated galaxy so we can compare the \eda~attenuation curves
to dust attenuation measured from observations. 
We begin with the well-established attenuation-slope relation: star-forming
galaxies with higher dust attenuation have shallower attenuation curves. 
This relation is a consequence of dust scattering dominating absorption at
low attenuation while dust absorption dominates at high
attenuation~\citep{gordon1994, witt2000, draine2003, chevallard2013}. 
In Figure~\ref{fig:slope}, we present the attenuation-slope relation of
star-forming galaxies with $\ssfr > 10^{-11}yr^{-1}$ using the
dust attenuation curves predicted by the \eda~for the median posteriors of
SIMBA (left), TNG (center) and EAGLE (right).
For comparison, we include the observed attenuation-slope relations of
GSWLC2 galaxies~\citep[grey shaded;][]{salim2020}.
For attenuation, we use $A_V$; for slope, we use the UV-optical slope, $S
= A(1500\AA)/A_V$. 
The contours mark the 68 and 95 percentiles. 

Most noticably, we find that the \eda~does not predict $A_V < 0.3$ for
star-forming galaxies.
This is a consequence of SIMBA, TNG, and EAGLE predicting star-forming galaxies
that are intrinsically more luminous than observations.  
All of the simulations have star-forming galaxies with intrinsic $M_r <
-21$ and $\gr < 0.5$ (Figure~\ref{fig:obs}). 
This is further corroborated by the $\sfr-M*$ relations in
Figure~\ref{fig:smf_msfr}, where the simulations all have star-forming
galaxies with $M_* > 10^{11}M_\odot$, not found in SDSS. 
To reproduce the SDSS optical color-magnitude relation these galaxies would
need to be significantly reddened and attenuated so any dust prescription would
need to assign high $A_V$ to star-forming galaxies.
Nevertheless, for $A_V > 0.3$, we find good agreement between the 
attenuation-slope relation predicted by the \eda~and observations. 
We refrain from further scrutinizing on $A_V$ values, since they can vary {\em
significantly} between different observational measurements --- even for the
same galaxy. 
SDSS star-forming galaxies, for instance, have significantly higher 
$A_V > 0.3$ according to the \cite{brinchmann2004} measurements
(Appendix~\ref{sec:slab}).


\begin{figure}
\begin{center}
    \includegraphics[width=0.5\textwidth]{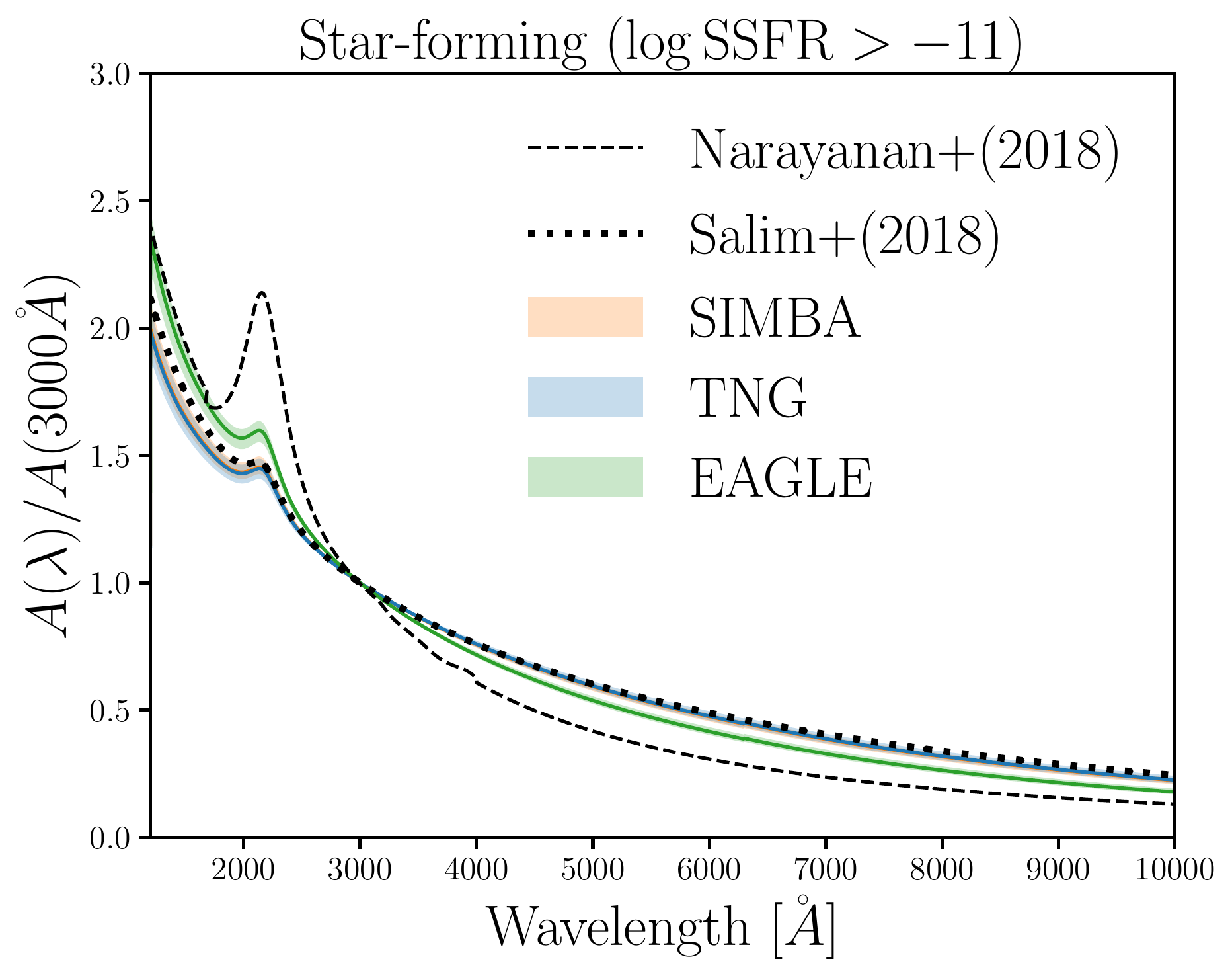}
    \caption{\label{fig:sfatten}
    The normalized attenuation curves of star-forming galaxies predicted by
    the \eda~for median posterior parameter values of SIMBA (orange), TNG
    (blue), and EAGLE (green).  
    We classify galaxies with $\log \ssfr > -11~yr^{-1}$ as star-forming. 
    The attenuation curves are normalized at $3000\AA$ and we mark the
    68 percentile of the attenuation curves with the shaded region.
    For comparison, we include $A(\lambda)/A(3000\AA)$ measurements from
    the~\cite{narayanan2018} radiative transfer simulation (dashed) and
    \cite{salim2018} observations (dotted).
    {\em The \eda~predict attenuation curves of star-forming galaxies
    are in good agreement with the attenuation curves measured from
    the simulation and observations in the literature.}
    }
\end{center}
\end{figure}

In addition to the attenuation-slope relation, we can also directly compare
the attenuation curves predicted by the \eda~to measurements from
observations for star-forming galaxies. 
In Figure~\ref{fig:sfatten}, we present the normalized attenuation curves
of star-forming galaxies predicted by the \eda~for the median posterior
parameter values of SIMBA(orange), TNG (blue), and EAGLE (green).
We again define galaxies with $\ssfr > 10^{-11}{yr}^{-1}$ as star-forming.
The attenuation curves are normalized at $3000\AA$ and we present the
variation in the attenuation curves in the shaded region, 68 percentile. 
For comparison, we include $A(\lambda)/A(3000\AA)$ from the
\cite{narayanan2018} radiative transfer simulation (dashed) and 
observations~\citep[][dotted]{salim2018}. 
The attenuation curve from \cite{salim2018} corresponds to star-forming
galaxies with $M_* > 10^{10.5}M_\odot$, a similar $M_*$ range as our
forward modeled samples. 
Since we do not vary the UV bump in our \eda~prescription, we ignore any
discrepancies in the amplitudes of the bump. 
\emph{Overall, we find good agreement between the \eda~attenuation curves for
star-forming galaxies and the attenuation curves from observations and
simulations.}

%% file: quiescent.tex
\begin{figure}
\begin{center}
    \includegraphics[width=0.9\textwidth]{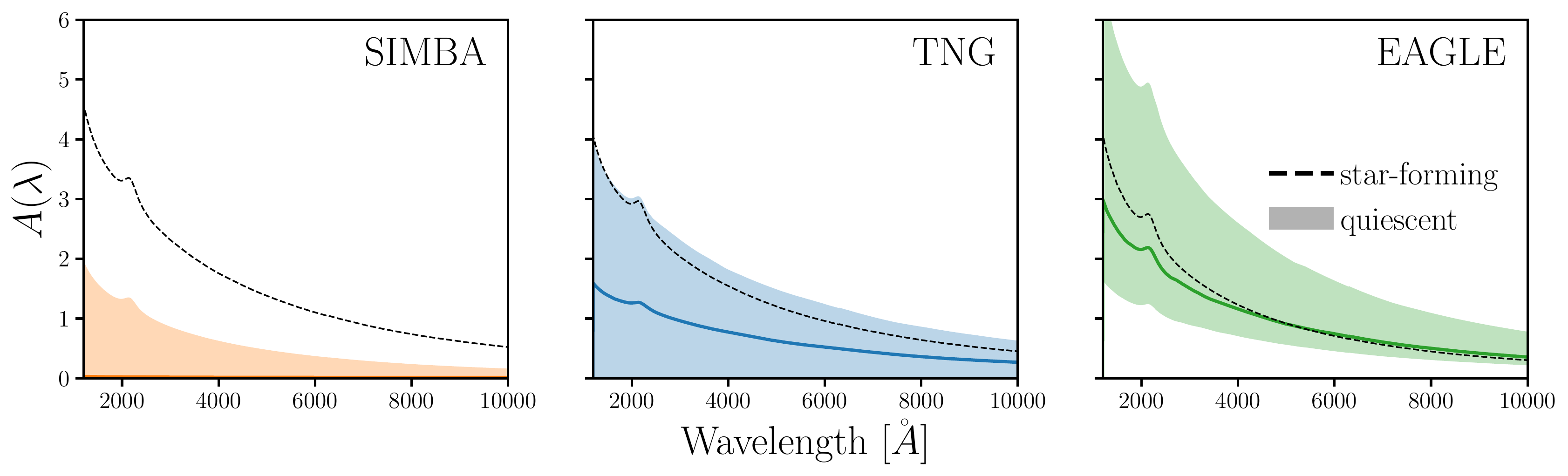}
    \caption{\label{fig:q_raw_atten}
    The attenuation curves of quiescent galaxies predicted by the \eda~for
    median posterior parameter values of SIMBA (left), TNG (center), and
    EAGLE (right).
    Galaxies with $\ssfr < 10^{-11}yr^{-1}$are classified as quiescent.
    We mark the 68 percentile of the attenuation curves with the shaded region
    and include the predicted attenuation curves of star-forming galaxies for
    comparison (dashed). 
    In all three simulations, \emph{the \eda~predicts attenuation curves that
    have lower amplitudes and shallower slopes than star-forming galaxies.}
    }
\end{center}
\end{figure}
\subsection{The Attenuation Curves of Quiescent Galaxies}  
In addition to star-forming galaxies, the \eda~also predicts dust attenuation
of quiescent galaxies. 
This is particularly valuable since dust attenuation in quiescent galaxies is
still poorly constrained by observations due to challenges in directly
measuring it from observations. 
For instance, methods that rely on IR luminosities can be contaminated by MIR
emission from AGN heating nearby dust~\citep{kirkpatrick2015}. 
SED fitting methods must also account for AGN MIR
emission~\citep{salim2016, leja2018, salim2018}. 
They also struggle to tightly constrain dust attenuation for quiescent galaxies
due to the degeneracies with star formation history and metallicity.
With a forward modeling approach, we circumvent these challenges. 
Instead, we derive the attenuation curves necessary for the simulated quiescent
population to reproduce the observed optical and UV photometry. 

In Figure~\ref{fig:q_raw_atten}, we present the attenuation curves of
quiescent galaxies predicted by the \eda~for the median posterior parameter
values of SIMBA (left), TNG (center), and EAGLE (right). 
Quiescent galaxies are classified using a $\ssfr < 10^{-11}{yr}^{-1}$ cut. 
Unlike Figure~\ref{fig:sfatten}, we do not normalize the attenuation curves at
$3000\AA$.  
For comparison, we include $A(\lambda)$ of star-forming galaxies predicted
by the \eda~for the corresponding simulation in each panel (dotted).
In all three simulations, quiescent galaxies have attenuation curves with
lower amplitudes and shallower slopes than star forming galaxies.

We predict $A(\lambda)$ with lower amplitude because quiescent galaxies in
SIMBA, TNG, and EAGLE are intrinsically only slightly more luminous than observations. 
In the top panels of Figure~\ref{fig:obs}, we see that the most luminous
galaxies with the highest $\gr$ color is $<0.5~mag$ brighter than the tip of
the red sequence in the SDSS color-magnitude relation. 
For SIMBA+\eda, where we predict $A(\lambda) \sim 0$, the most luminous and
optically red galaxies have comparable $M_r$ as the tip of the SDSS red sequence.
In contrast, the most luminous blue, star-forming, galaxies are $>1~mag$
brighter than the luminous end of the SDSS blue cloud.
Despite having lower attenuation than star-forming galaxies, in TNG and
EAGLE we predict significant dust attenuation in quiescent galaxies, $A_V > 0.5$.
Although this could be because TNG and EAGLE produce quiescent galaxies that
are intrinsically too luminous, the prescence of dust attenuation in quiescent
galaxies, which is typically neglected, has significant implications.  
For instance, it strengthens the evidence for the UV upturn phenomenon, the
unexpected detections of UV flux in quiescent galaxies~\citep[\eg][]{code1969,
oconnell1999, lecras2016, ali2018, dantas2021}. 
Constraints on the attenuation in quiescent galaxies may help discern
among the different hypotheses: residual star formation
activity~\citep[\eg~][]{kaviraj2007}, post-main-sequence stellar evolutionary
phases~\citep[\eg~][]{yi1997}, or binary systems~\citep[\eg~][]{han2007}.
Since the attenuation curves of quiescent galaxies are difficult to measure
from observations, the \eda~predictions highlight the advantages of
forward modeling and its complementarity with standard approaches. 

In Figure~\ref{fig:q_raw_atten}, we also find that quiescent galaxies have
shallower attenuation curves than star-forming galaxies. 
This is because SIMBA, TNG, and EAGLE all predict intrinsically UV-red galaxies
that do not require significant reddening. 
This is especially true for SIMBA and TNG, which predict significant
number of galaxies with intrinsic $\fnuv > 1$ (Figure~\ref{fig:obs}
and~\ref{fig:uv_sfh}).
These galaxies are quiescent ($\ssfr < 10^{-11}yr^{-1}$) and have high
$\fnuv$ due to a lack of recent star formation contributing to the
SED.
When we examine their SFHs, we find that, although they have more star
formation than the luminous UV-red galaxies discussed earlier, they have
little star formation in the last 1 Gyr. 
This implies that SIMBA, whose quiescent galaxies have the shallowest
attenuation curve, has the most efficient star-formation quenching among
the simulations. 

The mass resolution of the simulations can impact the SFHs of quiescent
galaxies and, thus, their observables.
The SFHs of simulated galaxies cannot include any star formation below the
resolution limit, which can affect the SEDs we compute from them. 
For recent star formation, this can have a significant impact on the SED,
especially in the FUV and NUV~\citep{leja2017}. 
SIMBA, TNG, and EAGLE have mass resolutions of $M_{\rm lim} = 1.82\times10^7M_\odot$, 
$1.4\times10^{6}M_\odot$, and $1.81\times10^6M_\odot$, respectively.
Even if we were to include a $< 100{\rm Myr}$ old stellar population
with total mass of $M_{\rm lim}$ in their SFHs, the impact on $\fnuv$ is small:
$\lesssim 0.1~mag$ for SIMBA and $\lesssim 0.01~mag$ for TNG and EAGLE.
Hence, mass resolution does not significantly impact the dust attenuation
we predict for quiescent galaxies.  

Despite the advantages of our forward modeling approach in deriving dust
attenuation curves for quiescent galaxies, we caution readers
that we only vary the \eda~parameters in this work.     
The \eda~predictions we present assume that the simulations accurately model
the star formation and metallicity histories of quiescent galaxies. 
Shortcomings in the galaxy formation models, and not the dust attenuation, may
be responsible for some of the differences between the simulations and
observations. 
For instance, we already find that quenching is too efficient in certain SIMBA
and TNG quiescent galaxies, which produce luminous UV-red galaxies not found in
SDSS (Section~\ref{sec:results}). 
TNG and EAGLE may also be producing quiescent galaxies that are
overall intrinsically too luminous, which then requires significant dust
attenuation to match observations.
In principle, we can vary both the \eda~parameters and the parameters
of the galaxy formation models and infer them simultaneously with a foward
modeling approach. 
We will explore this in future work.

%% file: galaxydust.tex
\begin{figure}
\begin{center}
    \includegraphics[width=0.9\textwidth]{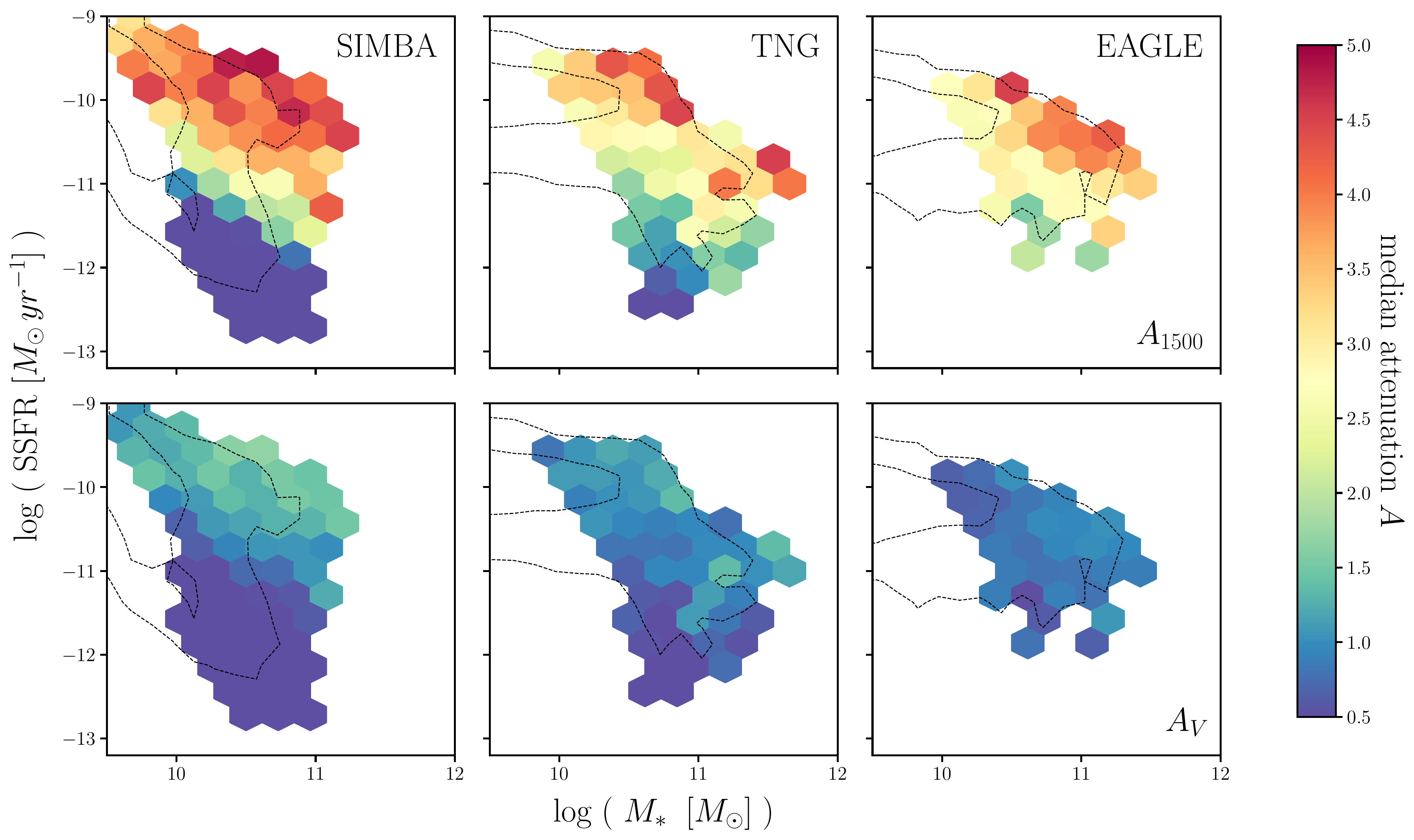}
    \caption{\label{fig:avmsfr}
        $M_*$ and $\ssfr$ dependence of dust attenuation at $1500 \AA$
        ($A_{1500}$; top) and at $5500\AA$ ($A_{V}$ bottom) predicted by the
        \eda~for SIMBA(left), TNG (center), and EAGLE (right). The colormap in each hexbin 
        represents the median attenuation for all simulated galaxies in the
        bin (right color bar). We only include bins with more than 10 galaxies.
        For reference, we include in each panel the $M_*-\ssfr$ relation of
        all galaxies from the simulations (black dashed).
        Overall, simulated galaxies with higher $M_*$ have higher dust
        attenuation at constant SSFR --- consistent with the literature.
        Furthermore, since previous works have primarily focused on star-forming
        galaxies, the \eda~provides new insight into the $\ssfr$ dependence of
        dust attenuation: simulated galaxies with higher $\ssfr$ have steeper
        attenuation curves. 
    }
\end{center}
\end{figure}

\subsection{The Galaxy -- Dust Connection}  
In this section, we examine the connection between dust attenuation and galaxy
properties in more detail. 
First, we examine the galaxy-dust connection using the $M_*$ and $\ssfr$
dependent parameterization of our
\eda~prescription (Eqs~\ref{eq:tauv} and~\ref{eq:delta}).
In Table~\ref{tab:posterior}, we list the median values and the 68\%
confidence interval of the inferred \eda~parameter posteriors for the 
three simulations. 
We find significant $\ssfr$ dependence, $m_{\tau,\ssfr} > 0$, consistent with
what we found in the previous section: quiescent galaxies have lower dust
attenuation than star-forming galaxies.
We also find significant $M_*$ dependence in $\tau_V$: $m_{\tau,M_*} > 0$.
$V$-band dust attenuation is higher for more massive galaxies.  
There is, however, little $M_*$ dependence in the slope of the dust
attenuation.

\begin{table}
    \caption{Inferred the Empirical Dust Attenuation Model Parameters}
    \begin{tabular}{lcccccc} \toprule
        & $m_{\tau,M_*}$ & $m_{\tau,\ssfr}$ & $c_\tau$ & $m_{\delta,M_*}$ & $m_{\delta,\ssfr}$ & $c_\delta$ \\[3pt] \hline\hline
        SIMBA   & $1.27\substack{+0.46\\-0.46}$ &
        $1.28\substack{+0.24\\-0.23}$ & $1.58\substack{+0.12\\-0.12}$ &
        $0.07 \substack{+0.12\\-0.11}$ & $0.13 \substack{+0.10\\-0.10}$ &
        $-0.18\substack{+0.04\\-0.04}$ \\
        TNG     & $0.57\substack{+0.44\\-0.53}$ &
        $0.62\substack{+0.21\\-0.20}$ & $1.34\substack{+0.19\\-0.21}$ &
        $-0.18\substack{+0.20\\-0.19}$ & $-0.19\substack{+0.15\\-0.16}$ &
        $-0.07\substack{+0.08\\-0.08}$ \\
        EAGLE   & $0.59\substack{+0.33\\-0.33}$ &
        $0.18\substack{+0.20\\-0.17}$ & $0.81\substack{+0.14\\-0.15}$ &
        $-0.13\substack{+0.17\\-0.18}$ & $-0.22\substack{+0.14\\-0.14}$ &
        $-0.34\substack{+0.08\\-0.08}$\\
        \hline
    \end{tabular} \label{tab:posterior}
\end{table}

We take a closer look at the $M_*$ and $\ssfr$ dependence of the attenuation
curve in Figure~\ref{fig:avmsfr}, where we present dust attenuation at
$1500\AA$ ($A_{1500}$; top) and $5500\AA$
($A_V$; bottom) as a function of $\log M_*$ and $\log \ssfr$ predicted by the
\eda~for SIMBA (left), TNG (center) and EAGLE (right). 
For each hexbin, the colormap represents the median attenuation for all
simulated galaxies in the bin. 
We only include bins with more than 10 galaxies. 
We include, for reference, the $M_* - \ssfr$ relation of all galaxies in the
simulations in black dashed contours, which mark the 68 and 95 percentiles.
We do not include a direct comparison with observations because there are large
variations between different measurements (Appendix~\ref{sec:slab}, see also
Figure~\ref{fig:av_obs}).  

In each panel, we find that SIMBA, TNG, and EAGLE galaxies with higher
$M_*$ have higher dust attenuation at constant SSFR --- consistent with the literature.
\cite{burgarella2005}, for instance, found significant positive $M_*$
dependence in $FUV$ attenuation in NUV-selected and FIR-selected samples. 
\cite{garn2010} and \cite{battisti2016} also found higher attenuation in
more massive SDSS star-forming galaxies. 
Most recently, \cite{salim2018} found higher $V$ and $FUV$ attenuation for
more massive star-forming galaxies in GSWLC2. 
For the $\ssfr$ dependence, we find that galaxies with higher $\ssfr$ have
higher $A_{1500}$ (top) and $A_V$ (bottom) as well as steeper slopes. 
The $\ssfr$ dependence is not as prominent in EAGLE (see also
Table~\ref{tab:posterior}), which has a narrower $\ssfr$ distribution than
SIMBA and TNG with no starburst galaxies or quiescent galaxies with $\ssfr <
10^{-12}yr^{-1}$. 
EAGLE has fewer intrinsically luminous star-forming galaxies
or UV-red galaxies (Figure~\ref{fig:obs}) and a narrower intrinsic $\gr$
and $\fnuv$ color distributions. 
To reproduce observations, it requires an overall attenuation and reddenning
without a significant $\ssfr$ dependence. 
Nevertheless, in all simulations, star-forming galaxies have slopes that
are consistent with observations (Section~\ref{sec:reproduce}) while
quiescent galaxies with the lowest $\ssfr$ have nearly flat attenuation
curves. 
In summary, we find that \emph{SIMBA, TNG, and EAGLE galaxies with higher
$M_*$ require overall higher dust attenuation and galaxies with higher
$\ssfr$ require steeper attenuation curves}.

%% file: discuss.tex
\subsection{Discussion}  
We make a number of assumptions and choices in our \eda~prescription. 
First, we use the slab model (Eq.~\ref{eq:slab}) to assign $A_V$ as a
function of $\tau_V(M_*, \ssfr)$ and randomly sampled $i$. 
This choice is motivated by the fact that the slab model reproduces
the correlation between attenuations and inclination found in star-forming
galaxies from observations~\citep{conroy2010b, wild2011, battisti2017,
salim2020} and simulations~\citep[\eg][]{chevallard2013,
narayanan2018, trayford2020}.
More importantly, we use the slab model because it is flexible enough to
reproduce a wide range of $A_V$ distributions. 
In Appendix~\ref{sec:slab}, we demonstrate that we can reproduce $p(A_V)$ of
SDSS star-forming galaxies as well as the GSWLC2 sample, which includes
quiescent galaxies.
We also use a parameterization of $\tau_V$ and $\delta$ that
depend linearly on $\log M_*$ and $\log {\rm SSFR}$. 
While the $M_*$ and $\ssfr$ dependence of $A_V$ is well-motivated and is
found in, for instance, the \cite{salim2018} GSWLC2 catalog (Appendix~\ref{sec:slab}), 
the linear dependence was chosen primarily for its simplicity.
The \eda~framework can be easily extended to more flexible
parameterizations. 
Though we already find good agreement with SDSS observations, a more
flexible parameterization can produce
even better agreement with the SDSS color-magnitude relations. 
The main challenges for a more flexible parameterization would be model
selection and finding a better motivated parameterization. 

We demonstrate in this work that accounting for dust attenuation is
essential when comparing simulations to observations. 
After all, none of the simulations reproduce the UV and optical
color-magnitude relation without dust (Figure~\ref{fig:obs}). 
Moreover, this work also demonstrates that our current lack of 
understanding of dust limits our ability to closely compare galaxy
formation models. 
Our \eda~prescription is built on what we currently know about dust attenuation
in galaxies: \eg~the \citealt{noll2009} parameterization, the UV bump, the
$M_*$, $\ssfr$ dependence, \emph{etc}.
Yet with the \eda, simulations that predict galaxy populations with
significantly different physical properties (Figure~\ref{fig:smf_msfr}) can
reproduce the same SDSS observations. 
For instance, SIMBA has significantly fewer massive galaxies above $M_* >
10^{11}M_\odot$ than TNG or EAGLE (see SMFs in Figure~\ref{fig:smf_msfr}). 
It also has $M_* < 10^{10}M_\odot$ starburst galaxies with $\ssfr >
10^{-9.5}yr^{-1}$ (see also \citealt{dave2019} Figures~5 and 6)
that are not found in TNG or EAGLE (Figure~\ref{fig:avmsfr}). 
Meanwhile, star-forming galaxies in TNG have significantly higher $\sfr$s than
in EAGLE.
Despite all these differences, by adjusting dust attenuation, the simulations
consistently reproduce the observations.

All this suggests that dust is highly degenerate with the differences between simulations. 
Put another way --- if we were to marginalize over dust in our comparison to
observations, we would not be able to differentiate between the different
galaxy physics prescriptions in the simulations. 
Hence, current limitations in our understanding of dust are a major bottleneck
for investigating galaxy formation with simulations.
In the next paper of the series, Starkenburg et al. (in preparation), we
will examine whether we can compare the prescriptions for star formation
quenching in different galaxy formation models once we include the
\eda~framework.


\begin{figure}
\begin{center}
    \includegraphics[width=0.45\textwidth]{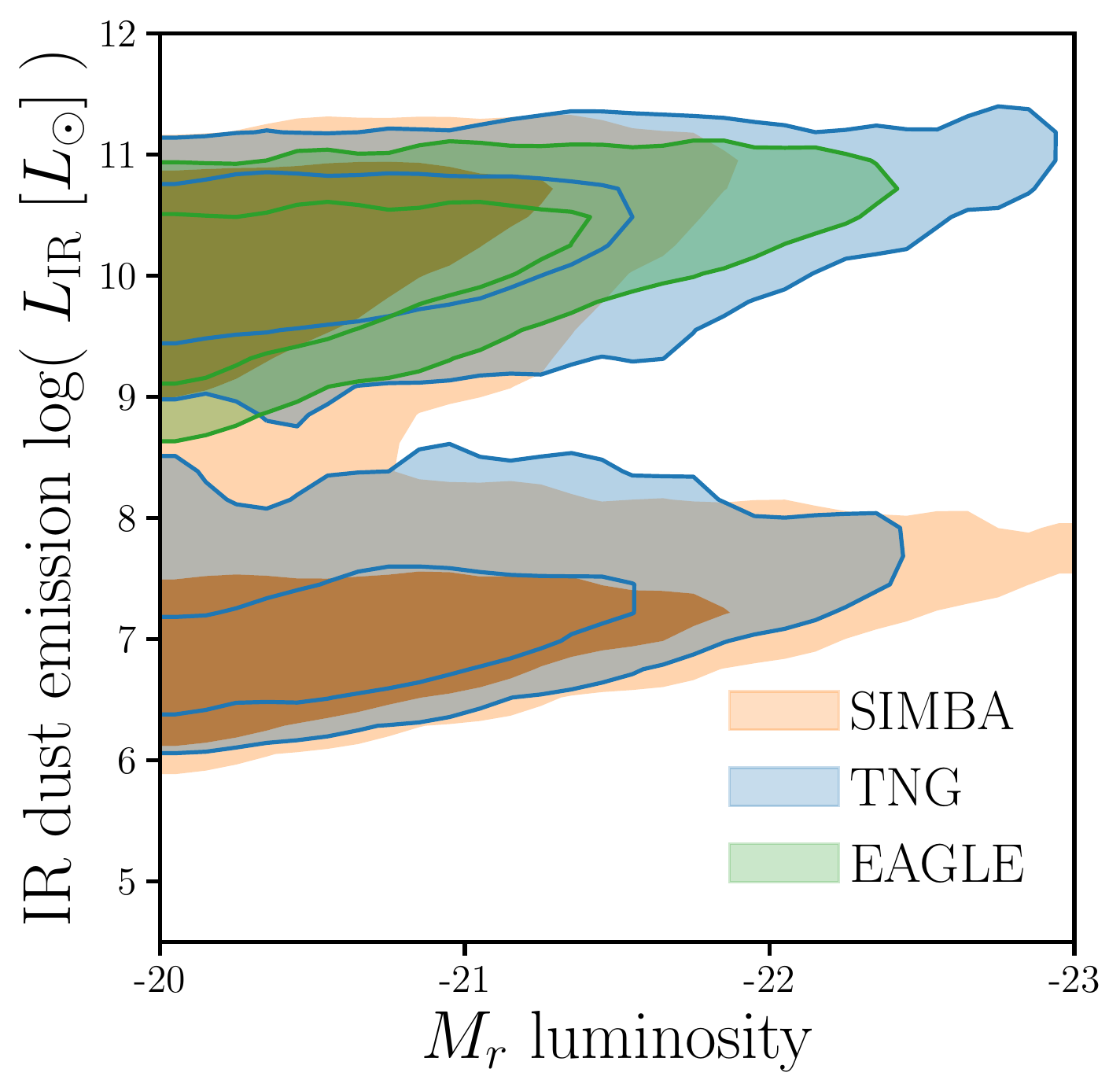}
    \caption{\label{fig:lir}
    IR dust emission luminosity predicted by the \eda~with median parameter
    values of the SIMBA (orange), TNG (blue), and EAGLE (green) posteriors as a
    function of $M_r$. 
    The dust emission is estimated assuming the \cite{dacunha2008} energy balance.
    Despite reproducing the same SDSS UV and optical color-magnitude relations,
    because the simulations require different amounts of dust attenuation to do
    this, \emph{the \eda~predicts significantly different IR dust emissions}.
    Therefore, including IR observations will significantly improve the
    constraints on \eda~parameters and allow us to better differentiate galaxy
    formation models.
    }
\end{center}
\end{figure}

Fortunately, there are many avenues for improving our understanding of dust
with a forward modeling approach. 
In this work, we used a restrictive SDSS galaxy sample with a $M_r < -20$,
$M_{FUV} < 13.5$, and $M_{NUV} < -14$ completeness limit. 
This selection excludes star-forming galaxies below $M_* \lesssim 10^{10}M_\odot$ and quiescent
galaxies below $M_* \lesssim 10^{10.5}M_\odot$ (Figure~\ref{fig:avmsfr}). 
Instead of imposing this completeness limit, we can include the actual SDSS
selection function in the forward 
model~\citep[\eg~][]{dickey2020}. 
This would allow us to compare the simulations with \eda~to the entire SDSS
sample, a substantially larger sample with a wider range of galaxies. 
Upcoming surveys, such as the Bright Galaxy Survey (BGS) of the Dark Energy
Spectroscopic Instrument~\citep[DESI;][]{desicollaboration2016, ruiz-macias2020} 
and the galaxy evolution survey of the Prime Focus
Spectrograph~\citep[PFS;][]{takada2014,tamura2016}, will vastly expand galaxy
observations. 
BGS, for instance, will measure $10\times$ the number of galaxy spectra as
SDSS out to $z\sim0.4$  and with its $r\sim20$ magnitude limit will probe
a broader range of galaxies.
Such observations will allow us to place tighter constraints on
the \eda~parameters, which may enable comparisons of the underlying galaxy
formation models and shed light on dust in a broader range of galaxies.
    
In this work, we only use observables derived from UV and optical
photometry, which means that we have only examined one side of the impact
that dust has on galaxy spectra.
While dust attenuates light in the optical and UV, it emits light in
IR. In fact, even though the simulations reproduce the same SDSS UV and
optical color-magnitude relations with the \eda, they predict significantly 
different dust emission in the IR. 
In Figure~\ref{fig:lir}, we present IR dust emission luminosity, $L_{\rm
IR}$, predicted by the \eda~with median parameter values of the SIMBA 
(orange), TNG (blue), and EAGLE (green) posteriors as a function of the
$r$-band absolute magnitude, $M_r$.
The dust emissions are estimated using the standard energy balance assumption
--- \ie~all starlight attenuated by dust is reemitted in the IR~\citep{dacunha2008}. 
Most noticably, SIMBA and TNG have bimodal distributions of dust emission
while EAGLE only has luminous IR dust emissions. 
This is because EAGLE requires significant dust attenuation in all
galaxies while SIMBA and TNG require quiescent galaxies to have
substantially lower dust attenuation than star-forming galaxies
(Figure~\ref{fig:q_raw_atten}). 
The luminous mode of the $L_{\rm IR}$ distributions, however, are in good
agreement for all simulations, since they all predict comparable dust
attenuation in star-forming galaxies
(Figures~\ref{fig:slope} and~\ref{fig:q_raw_atten}). 
When we compare the IR dust emission of SIMBA+\eda~and TNG+\eda~further, we
find that TNG+\eda~produces more luminous galaxies with high IR dust
emission ($M_r < -22$ and $L_{\rm IR} > 10^9L_\odot$) since it has more
intrinsically luminous star-forming galaxies~(Figure~\ref{fig:obs}).
On the other hand SIMBA+\eda~has more luminous galaxies with fainter IR
dust emissions, which correspond to the anomolous luminous UV-red galaxies
highlighted in Figure~\ref{fig:uv_sfh}.

While dust attenuation can be adjusted to reproduce UV and optical
observations, since IR dust emission measures the total attenuation, IR
observations can place a limit on the total impact of dust and thereby
break the degeneracies between dust and the galaxy physics in simulations.
While some upcoming surveys, such as BGS, will have existing near-IR
photometry from NEOWISE~\citep{meisner2018}, future observations will
dramatically expand the information we have in IR.
\emph{Nancy Grace Roman Space Telescope} and \emph{James Webb Space
Telescope}, for instance, will provide valuable near and
mid-IR observations. 
Meanwhile, IR observations at even longer wavelengths will come from
Atacama Large Millimeter/submillimeter Array or future facilities
such as the Next-Generation Very Large Array and \emph{Origins Space Telescope}.

%% file: summary.tex
\section{Summary}
In this work, we present the \eda, a framework for applying dust attenuation to
simulated galaxy populations. It uses a parameterization of 
the attenuation curves motivated from observations~\citep{noll2009} and
assigns attenuation curves to simulated galaxies based on their physical properties ($M_*$ and SSFR).
We apply the \eda~to 
three state-of-the-art hydrodynamical simulations (SIMBA, TNG, and EAGLE) and
forward model the optical and UV color-magnitude relations. We then compare
the forward modeled simulations to a $M_r < -20, M_{FUV} < -13.5, M_{NUV} <
-14.0$ complete SDSS galaxy sample using likelihood-free inference. Based on
this comparison, we find the following results: 

\begin{itemize}
    \item Dust attenuation is essential for our hydrodynamical simulations to
        reproduce observations.
        Without dust, SIMBA, TNG, and EAGLE all struggle to reproduce the
        observed UV and optical color-magnitude relation. 
        They predict galaxies that are intrinsically much bluer and more
        luminous in the optical and with broader distributions of $\fnuv$
        colors than SDSS. 
    \item With the \eda, each of the simulations are able to produce UV and
        optical color-magnitude relations in good agreement with SDSS
        observations. 
        However, we find that star formation quenching is too efficient for
        some of the most massive quiescent galaxies in SIMBA and TNG. 
        When forward modeled, these galaxies appear as luminous UV-red
        galaxies outside the observed UV color-magnitude relation that 
        cannot be reconciled by dust. 
    \item The attenuation curves of star-forming galaxies predicted by the
        \eda~for the simulations are in good agreement with the observed
        attenuation-slope relation. 
        They also closely reproduce the observed attenuation curves
        of star-forming galaxies. 
        The simulations, however, predict star-forming galaxies that are
        intrinsically more luminous than observations. Hence, the \eda~does
        not predict star-forming galaxies with $A_V < 0.3$, found in the
        literature. 
    \item Lastly, we predict the attuenation curves of
        quiescent galaxies, which observations struggle to directly
        measure.
        We find that quiescent galaxies have $A(\lambda)$ curves
        with lower amplitudes and shallower slopes than star-forming
        galaxies. 
        Over the entire population, we find that more massive galaxies have
        higher overall dust attenuation while galaxies with higher SSFR
        have steeper attenuation curves. 
\end{itemize}

Our results clearly demonstrate that the \eda~and a forward modeling approach
provides key insights into dust attenuation. For those uninterested in dust,
the \eda~also provides a computationally feasible framework for marginalizing
over dust when comparing simulations to observations. 
However, we find that dust attenuation is {\em highly} degenerate with
differences in their galaxy physics prescriptions.
Even though the simulations predict galaxy populations with significantly
different physical properties, there is enough uncertainty in our
understanding of dust that by adjusting attenuation they can all
reproduce the same SDSS observations.
This also suggests that any comparisons across simulations must marginalize
over dust attenuation or run the risk of overinterpretation. 
Therefore, our current understanding of dust, or lack of, limit our ability to
distinguish between the various hydrodynamical models and is a major bottleneck
for investigating galaxy formation using simulations.

The forward modeling approach we present offers many avenues for improving on
our understanding of dust. In this paper, we used a restrictive observational
galaxy sample from SDSS. Comparison to a larger observed sample will place
tighter constraints on \eda~parameters and enable better
differentiation between the simulations. One way to expand the observed galaxy
sample would be to remove the completeness limits and include the SDSS survey
selection in our forward model. Upcoming surveys, such as the DESI
Bright Galaxy Survey and the PFS Galaxy Evolution Survey, will also soon
provide much larger observational galaxy samples. Furthermore, IR observations,
which measure dust emission and trace the total dust attenuation, have the
potential to tightly constrain the \eda~parameters and therefore break degeneracies
between dust and the galaxy physics in simulations. 
In the next paper of the series, we will use the forward modeling approach
with the \eda~to investigate star formation quenching in galaxy formation
models. 
In other future works, we will apply the \eda~and a forward modeling approach
to more statistically powerful samples and include IR observables in order to
tightly constrain and reveal new insights into dust attenuation.

%% file: slab.tex
\begin{figure}
    \begin{center}
        \includegraphics[width=0.66\textwidth]{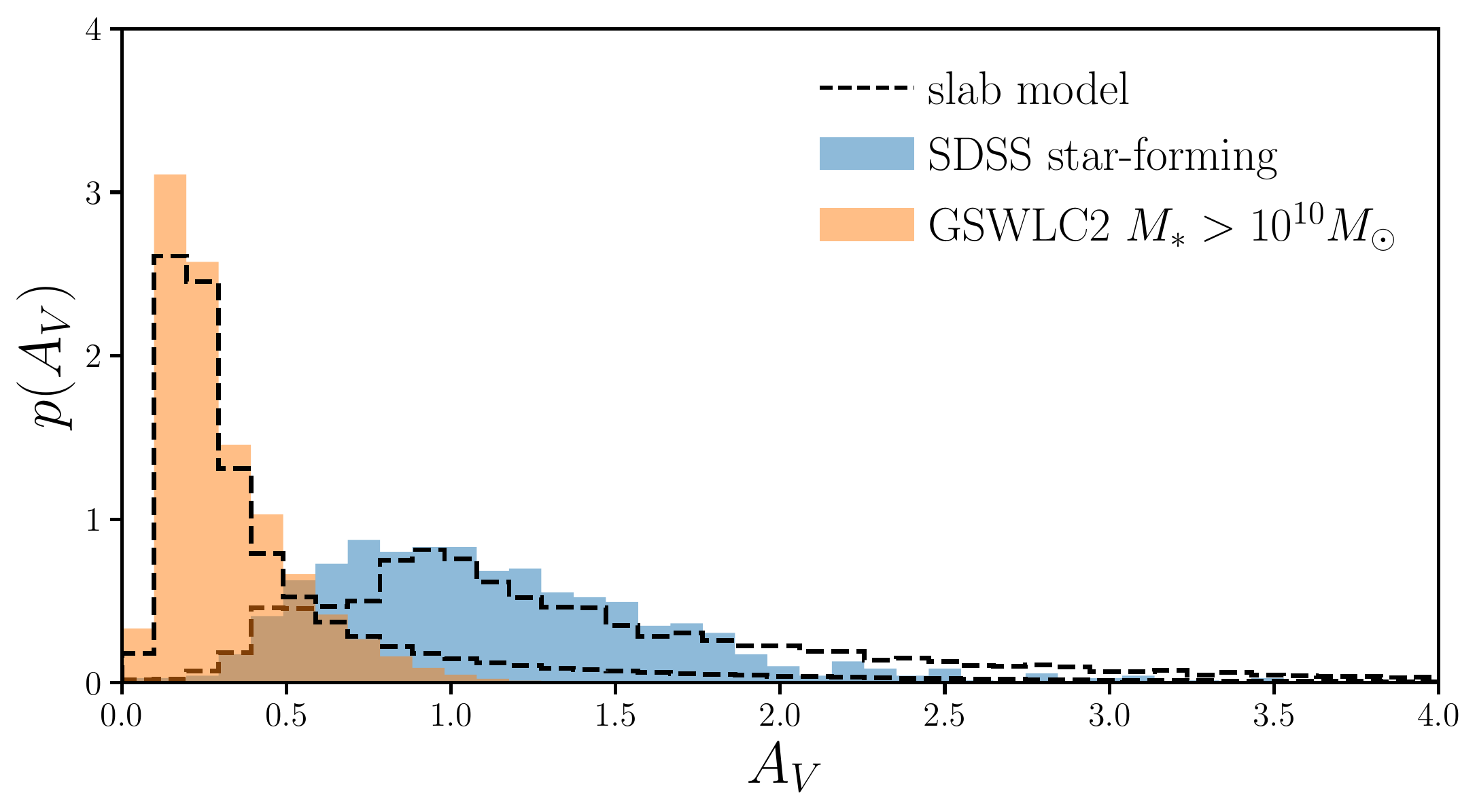} 
        \caption{\label{fig:av_dist}
        The $A_V$ distributions, $p(A_V)$, generated from the slab model (Eq.~\ref{eq:slab};
        black dash) compared to $p(A_V)$ of star-forming galaxies our SDSS
        sample (blue; Section~\ref{sec:obs}) and of $M_* > 10^{10}M_\odot$
        star-forming and quiescent galaxies in the GSWLC2 sample (orange). 
        The $A_V$ values for both observations are derived using SED
        fitting but with different bands and methodologies~\citep{brinchmann2004, salim2018}. 
        For the slab model, we generate $A_V$ values for each galaxy
        in the SDSS and GSWLC2 samples using Eq.~\ref{eq:slab} with its
        measured $M_*$ and $\ssfr$ and randomly sampled $i$. 
        Despite the significant differences between the $p(A_V)$ of SDSS and
        GSWLC2, the slab model is able to generate $p(A_V)$ in good agreement
        with both observations using parameter values within the
        Table~\ref{tab:free_param} prior range. 
        Therefore, the slab model provides a sufficiently flexible prescription
        for our \eda.
        }
    \end{center}
\end{figure}

\section{The Slab Model Based EDA}  \label{sec:slab} 
In our \eda~prescription, we use the slab model to determine $A_V$, the
amplitude of attenuation, as a function of a randomly sampled inclination,
$i$, and $\tau_V$ (see Eq.~\ref{eq:slab} in Section~\ref{sec:dem}).
The slab model is based on the assumption that dust in galaxies have
slab-like geometry and are illuminated by the stellar radiation
source~\citep{somerville1999}. 
For a given $\tau_V$, the attenuation depends solely on the orientation of the galaxy. 
While this simplification reproduces the correlation between $A_V$ and $i$
found in observed star-forming galaxies~\citep[\eg][]{conroy2010, wild2011,
battisti2017, salim2020}, it ignores the detailed star-to-dust geometry that
impacts the attenuation curve. 
It also does not provide a physically-motivated prescription for quiescent
galaxies, which typically have elliptical morphologies.
Despite its limitations, the slab model provides a robust empirical
prescription that allows us to produce realistic distributions of $A_V$. 

In Figure~\ref{fig:av_dist}, we compare the $A_V$ distributions, $p(A_V)$,
of star-forming galaxies in SDSS (blue) and galaxies in the
\cite{salim2018} GSWLC2 sample (orange) to $p(A_V)$ generated from the
slab model (black dashed). 
The $A_V$ values of the SDSS are derived using SED fitting from the
\cite{brinchmann2004} MPA-JHU catalog.
The GSWLC2 $A_V$ values are also derived from SED fitting UV and optical
photometry from GALEX and SDSS observations as well as mid-IR photometry from WISE. 
The GSWLC2 $p(A_V)$ includes all galaxies, including quiescent ones, above
$M_* > 10^{10}M_\odot$. 
We generate two $p(A_V)$ with the slab model for the SDSS and GSWLC2
samples separately.
For each SDSS/GSWLC2 galaxy, we determine $A_V$ by uniformly sampling 
$\cos i$ from 0 to 1 and derive $\tau_V$ (Eq.~\ref{eq:tauv}) with the
galaxy's measured $M_*$ and $\ssfr$. 
We pick $\mtaum, \mtaus, c_\tau$ values within the prior range
(Table~\ref{tab:free_param}) by hand to roughly reproduce the SDSS and GSWLC2 $p(A_V)$
distributions. 

Galaxies in SDSS and GSWLC2 have substantially different $p(A_V)$. 
While the galaxy populations only partially overlap, this difference is 
{\em mostly} due to inconsistencies in the $A_V$ measurements of MPA-JHU and GSWLC2 --- even
for the same galaxy.
This difference in $p(A_V)$ illustrates the challenges in directly measuring
dust attenuation. 
Despite the dramatic differences between the two, the slab model can
produce $p(A_V)$ in good agreement with both observed distributions. 
We therefore conclude that the slab model provides a sufficiently flexible
prescription to sample a realistic distribution of $A_V$. 

\begin{figure}
\begin{center}
    \includegraphics[width=0.8\textwidth]{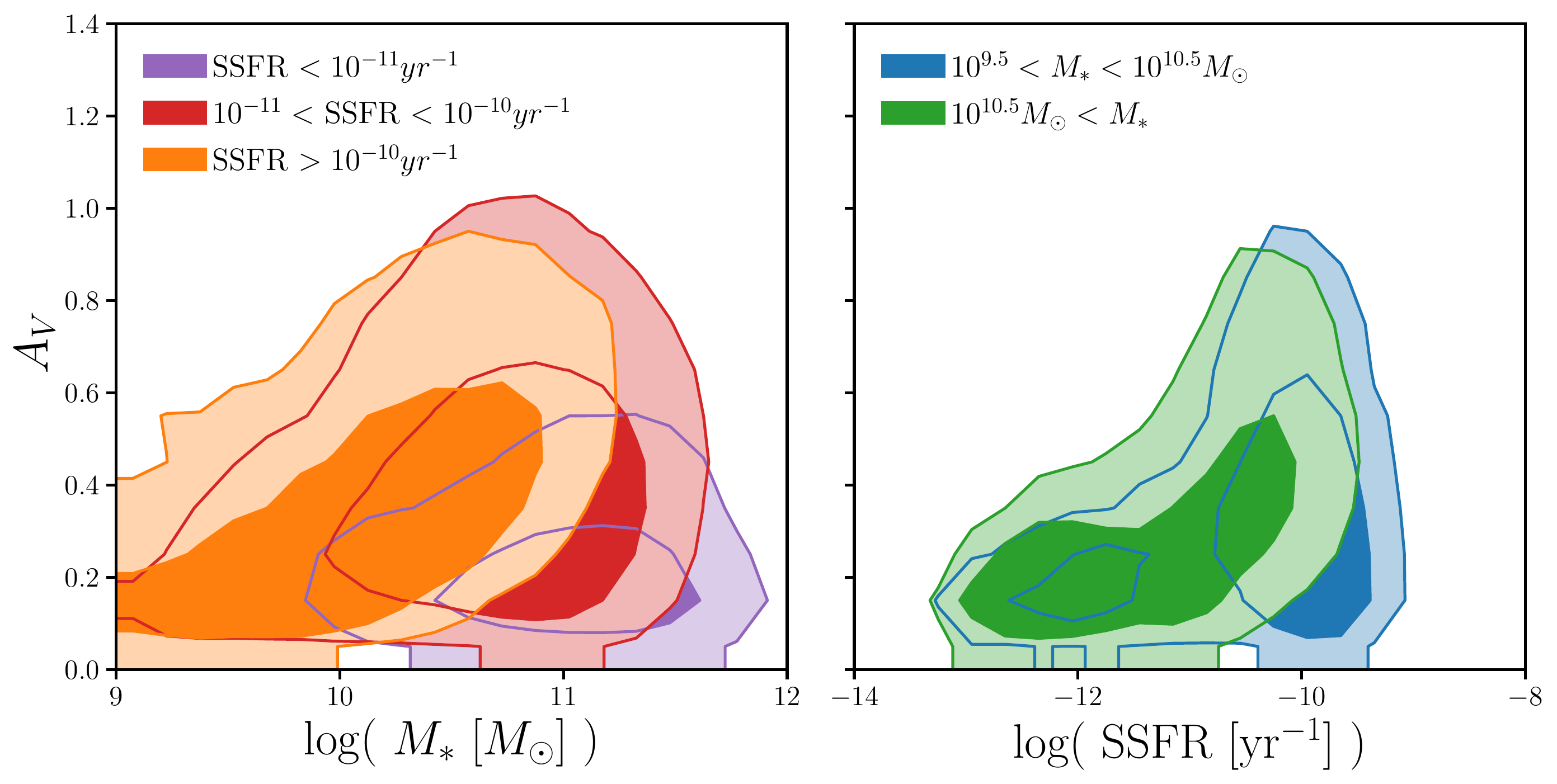}
    \caption{\label{fig:dep}
    Dependence of $A_V$ on $M_*$ (left) and $\ssfr$ (right) for the
    \cite{salim2018} GSWLC2 sample.
    In the left panel, we divide the GSWLC2 sample into bins of $\ssfr$: 
    $\ssfr < 10^{-11}yr^{-1}$ (purple), 
    $10^{-11} < \ssfr < 10^{-10}yr^{-1}$ (red),
    and  $10^{-10} < \ssfr$ (orange). 
    In each of the $\ssfr$ bins, we find significant $M_*$ dependence. 
    In the right panel, we divide the sample into bins of $M_*$:  
    $10^{9.5} < M_* < 10^{10.5}M_\odot$ (blue) and $10^{10.5} M_\odot < M_*$ (green).
    In the $M_* > 10^{10.5}M_\odot$ bin, which roughly corresponds to our
    SDSS sample, we find significant $\ssfr$ dependence.
    The $M_*$ and $\ssfr$ dependence in $A_V$ we find in GSWLC2 is
    consistent with previous works and provides further motivation for our
    \eda~prescription.
    }
\end{center}
\end{figure}

In addition to the slab model, in the \eda, we also use a linear
dependence on $M_*$ and $\ssfr$ in the $V$ band optical depth,
$\tau_V$ (see Eq.~\ref{eq:tauv}).
This parameterization is motivated by observations that find significant
correlation between $A_V$ and $M_*$ and $\ssfr$~\citep[\eg~][]{garn2010, battisti2016, salim2020}. 
We take a closer look at this correlation using the GWSLC2 sample in
Figure~\ref{fig:dep}.
We present the dependence of $A_V$ on $M_*$ (left panel) and $\ssfr$ (right
panel). 
In the left panel, we divide the GSWLC2 galaxies by $\ssfr$: 
$\ssfr < 10^{-11}yr^{-1}$ (purple), $10^{-11} < \ssfr < 10^{-10}yr^{-1}$
(red), and  $10^{-10} < \ssfr$ (orange). 
For each of the $\ssfr$ bins, we find significant $M_*$ dependence in
$A_V$: more massive galaxies have higher $A_V$.
In the right panel, we divide the galaxies by $M_*$: 
$10^{9.5} < M_* < 10^{10.5}M_\odot$ (blue) and $10^{10.5} M_\odot < M_*$
(green).
In both $M_*$ bins, galaxies with higher $\ssfr$ have higher $A_V$. 
The dependence is stronger stronger for galaxies with $M_* > 10^{10.5}M_\odot$, 
which roughly corresponds $M_*$ limit of our forward model (see
Figure~\ref{fig:avmsfr}). 
Overall, the $M_*$ and $\ssfr$ dependence we find in $A_V$ from the GSWLC2 sample is
consistent with previous observations and further motivates our
\eda~prescription.

\begin{figure}
\begin{center}
    \includegraphics[width=\textwidth]{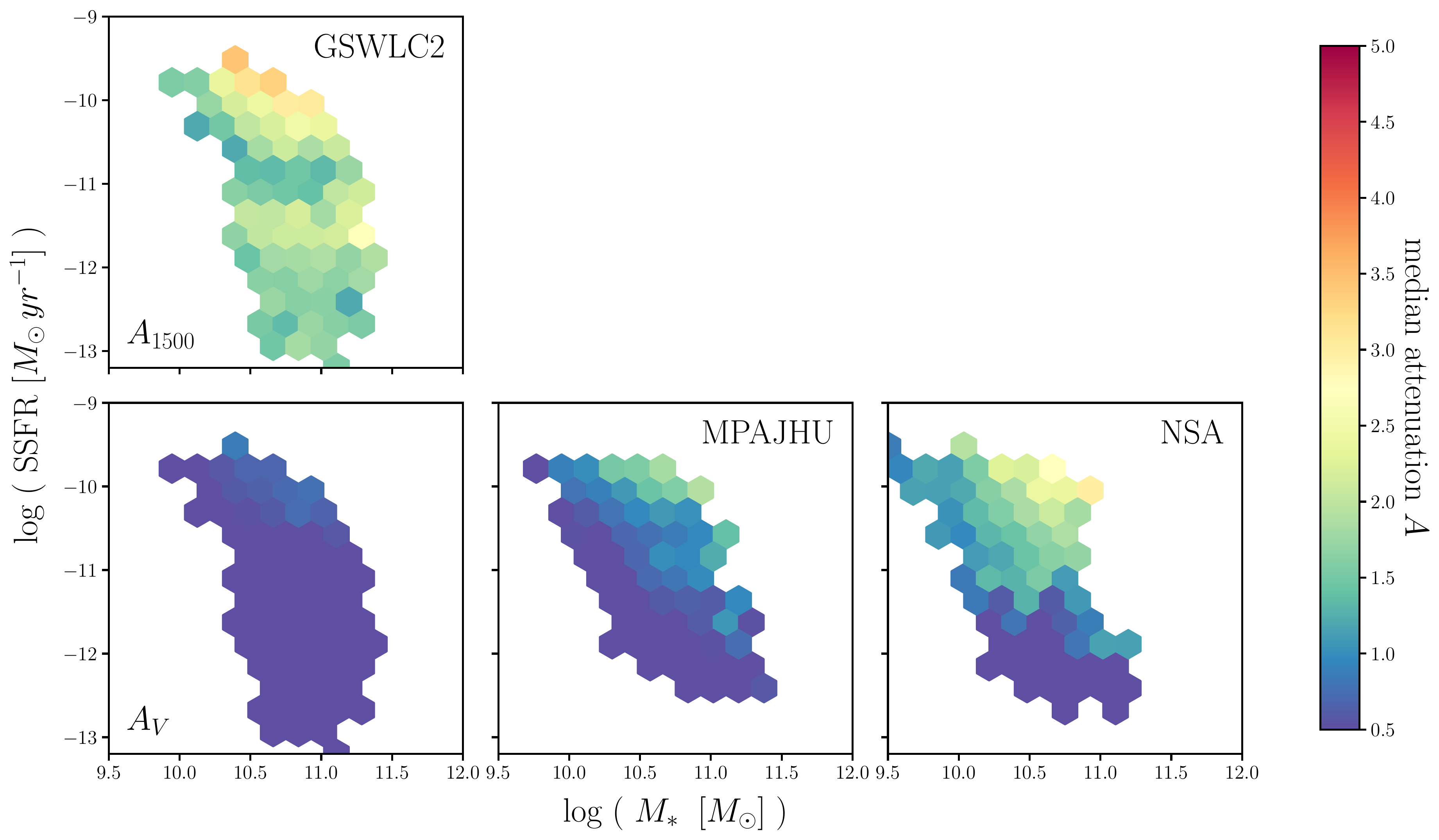}
    \caption{\label{fig:av_obs}
    $M_*$ and $\ssfr$ dependence of dust attenuation at $1500 \AA$
    ($A_{1500}$; top) and at $5500\AA$ ($A_{V}$; bottom) of SDSS galaxies. 
    The sample includes 2361 galaxies that pass our selection cut
    (Section~\ref{sec:obs}) and are also part of the GSWLC2 and MPA-JHU
    samples. 
    In the left panels, we use $A_{1500}$ and $A_V$ from GSWLC2.
    In the center panel, we use $A_V$ from MPA-JHU. 
    In the right panel, we use $A_V$ from the NSA. 
    The $M_*$ and $\ssfr$ in each panel are from the respective samples. 
    Same as Figure~\ref{fig:avmsfr}, the colormap in each hexbin represents the
    median attenuation for all galaxies in the bin (color bar).
    We only include bins with more than 5 galaxies.
    The bottom panels illustrate that $A_V$ measurements from GSWLC2, MPA-JHU,
    and NSA differ {\em significantly} even for the same galaxies.
    We, therefore, do not directly compare our \eda~predictions to observations. 
    }
\end{center}
\end{figure}

In Figure~\ref{fig:av_obs}, we present the $M_*$ and $\ssfr$ dependence of dust
attenuation in SDSS galaxies, which contains 2361 galaxies that pass our
selection cut and are also in the GSWLC2 and MPA-JHU samples.
In the top panel, we present $A_{1500}$ from GSWLC2 as a function of $M_*$ and
$\ssfr$.
In the bottom panels, we present $A_V$ from GSWLC2 (left), MPA-JHU (center),
and NSA (right).
The NSA $A_V$ measurements are derived assuming intrinsic Balmer decrement of
2.85, $R_V=3.1$ and \cite{odonnell1994} extinction. 
The colormap in each hexbin represents the median attenuation for galaxies in
the bin, same  as in Figure~\ref{fig:avmsfr}. 
Bins with less than 5 galaxies are omitted. 
For each observational sample (column), we use $M_*$ and $\ssfr$ from the
respective samples for consistency.
We find the same $M_*$ and $\ssfr$ dependence of $A_V$ as Figure~\ref{fig:dep}
even after our selection cut (bottom left). 
The bottom panels highlight that evven for the same galaxies, $A_V$ from
GSWLC2, MPA-JHU, and NSA have significant differently amplitudes. 
$M_*$ and $\ssfr$ are also significantly different across the samples.
Since observations have large discrepancies among dust attenuation measurements
and a detailed comparison is beyond the scope of this work, we refrain from
comparing our \eda~predicted dust attenuation (Section~\ref{sec:results}) to
observations.